\documentclass{elsart}

\usepackage{epsfig}

\usepackage{amssymb}

\begin{document}

\begin{frontmatter}

% Title, authors and addresses

% use the thanksref command within \title, \author or \address for footnotes;
% use the corauthref command within \author for corresponding author footnotes;
% use the ead command for the email address,
% and the form \ead[url] for the home page:
% \title{Title\thanksref{label1}}
% \thanks[label1]{}
% \author{Name\corauthref{cor1}\thanksref{label2}}
% \ead{email address}
% \ead[url]{home page}
% \thanks[label2]{}
% \corauth[cor1]{}
% \address{Address\thanksref{label3}}
% \thanks[label3]{}

\title{Double-component convection due to different boundary
conditions in an infinite slot diversely oriented to the
gravity\thanksref{*}}

% use optional labels to link authors explicitly to addresses:
% \author[label1,label2]{}
% \address[label1]{}
% \address[label2]{}

\author{N.~Tsitverblit\thanksref{**}}

\address{Department of Fluid Mechanics and Heat Transfer, 
Tel-Aviv University, Ramat-Aviv 69978, Israel}
\thanks[*]{NOTICE: this is the author's version of a work that was accepted for publication in
Annals of Physics. Changes resulting from the publishing process, such as peer review, editing,
corrections, structural formatting, and other quality control mechanisms may not be reflected
in this document. Changes may have been made to this work since it was submitted for publication.
A definitive version was subsequently published in Annals of Physics
[Ann. Phys. (2007) 322(8) 1727-1770]\newline DOI 10.1016/j.aop.2006.10.001.}
\thanks[**]{Address for correspondence: 
1 Yanosh Korchak Street, apt. 6, Netanya 42495, Israel; 
e-mail: naftali@eng.tau.ac.il}

\begin{abstract}
% Text of abstract
Onset of small-amplitude oscillatory and both small- and finite-amplitude steady
double-component convection arising due to component different boundary conditions in an infinite
slot is studied for various slot orientations to the gravity. The main focus is on two compensating
background gradients of the components. The physical mechanisms underlying steady and oscillatory
convection are analyzed from the perspective of a universally consistent understanding of the effects
of different boundary conditions. In a horizontal slot with inviscid fluid addressed by \mbox{Welander
[Tellus, Ser. A 41 (1989) 66]}, oscillatory convection sets in with the most unstable wave number
and oscillation frequency being zero. Exact expressions for the critical fixed-value background
gradient and the respective group velocity at zero wave number are derived from the long-wavelength expansion
both for the horizontal slot with independently varying background gradients and for the inclined slot with
the compensating gradients. In the horizontal slot with viscous fluid, the dissipation of along-slot
perturbation-cell motion reduces efficiency of the oscillatory instability feedback and thus prevents
the most unstable wavelength from being infinite. Based on this interpretation, the oscillatory instability
of a three-dimensional (3D) nature is predicted for an interval of long two-dimensional (2D) wavelengths
in an inclined slot, and such 3D instability is indeed shown to arise. Related general conditions for
three-dimensionality of most unstable disturbances are also formulated. As the slot orientation changes from the
horizontal by angle $\theta$ ( $\geq\pi/2$), the oscillatory 2D marginal-stability boundaries in inviscid and viscous
fluid are expected to eventually transform into respective steady ones. Oscillatory instability in the vertical
slot with viscous fluid, first reported by \mbox{Tsitverblit [Phys. Rev. E 62 (2000) R7591]}, is of a
quasisteady nature. Its (new) mechanism is identified. It is underlain by differential gradient diffusion. As
the horizontal slot at $\theta=\pi$, addressed by \mbox{Tsitverblit [Phys. Fluids 9 (1997) 2458]},
changes its orientation to vertical, the wave number interval of linear steady instability shrinks to the
vicinity of the most unstable zero wave number and vanishes. Consistently with the basic nature of
finite-amplitude steady convection being the same in the horizontal and vertical slots, the
respective convective flows are continuously transformed into each other. The dissimilarity
between the nature of finite-amplitude steady convective flows in the horizontal slot with
$\theta=0$, revealed by \mbox{Tsitverblit [Phys. Lett. A 329 (2004) 445]}, and
that in the vertical slot is shown to eventually give rise to a
region of hysteresis in $\theta\in(0,\pi/2)$.
\end{abstract}

\end{frontmatter}

% main text
%\section{}
%\label{}

% The Appendices part is started with the command \appendix;
% appendix sections are then done as normal sections
% \appendix

% \section{}
% \label{}
\section{\label{s:i}Introduction}
This work, some of whose aspects were promulgated
in \cite{r:tegs}, addresses onset of double-component,
buoyancy-driven convection resulting from the boundary
conditions for one component being different from those for
the other. Such convection has recently been identified as a
fundamentally new class of hydrodynamic instabilities underlying
the formation of spatial and temporal flow patterns from a steady
equilibrium state of spatially homogeneous fluid. The main emphasis
of this study is on developing a universally consistent perspective
for understanding the effects of different orientation
of the fluid domain to the gravity. 

Double-component convection is relevant to the phenomena in small-scale
oceanography \cite{r:scm}, geology \cite{r:hsb}, geodynamo \cite{r:bus}, and
such areas of astrophysics as ordinary evolution of stars \cite{r:spghpc} and
core-collapse supernova explosions \cite{r:bm}. It has technological applications as well,
in particular to crystal growth \cite{r:crsk}. Double-component convection has also been
suggested as a possible cause of layering phenomena in colloidal suspensions \cite{r:mceg}
and as the origin of pattern formation in soap films \cite{r:mw}. Early reviews of some
of the above applications of the effects of different component diffusivities in
double-component convection can be found in \cite{r:tur}.

In addition, convective flows are commonly used in fundamental
studies of transition to turbulence and nonlinear pattern formation
\cite{r:br,r:ch}. Double-component flows where a distinction between the
components comes from component different boundary conditions are also of basic
significance in the context of their applications to large-scale environmental
and turbulent processes. In particular, such processes are \mbox{Langmuir}
circulations \cite{r:l83th04} and the global ocean thermohaline circulation
\cite{r:stmw,r:qgdm}. Except for limited aspects of the latter context,
however, two-component convection in pure fluid has been previously
addressed mostly in the framework of only the effects of different
diffusion coefficients \cite{r:stnf,r:ver,r:stnl}.

\mbox{Welander~\cite{r:wel}} was the first to highlight the onset
of oscillatory convection resulting from an unequal effect of different
boundary conditions on the component diffusion gradients in perturbed
state. Such convection was demonstrated in \cite{r:wel} to arise in a
horizontal layer of inviscid fluid with a double-component, statically
stable net stratification. The instability mechanism described in
\cite{r:wel} has later been recognized by the present author \cite{r:twh,r:tbc}
as being conceptually analogous to that driving oscillatory convection in
the diffusive regime of the classical double-diffusion \cite{r:stnf,r:ver}.
It has also been suggested in \cite{r:twh,r:tbc,r:tlh} that such analogy
between the effect of different boundary conditions and that of unequal
diffusivities is of a generic nature. Near a boundary towards which
across-slot perturbation motion is directed, the component with its
boundary value fixed would have a higher perturbation gradient than
the component whose boundary condition is specified in terms of the
flux. The different rates of gradient diffusion resulting from such
a disparity could thus trigger double-component convection analogously
to the effect of unequal component diffusivities in
conventional double-diffusive convection. 

In terms of the analogy just described, the mechanism
of steady convection under the stratification inverse to
that in \cite{r:wel} has been shown in \cite{r:twh,r:tbc}
to be conceptually reminiscent of the finger instability in the
classical double-diffusion \cite{r:stnf}. (The mathematical problem 
discussed in \cite{r:twh,r:tbc} also happens to describe double-component
\mbox{Langmuir} circulations \cite{r:l83th04}.) In application to two
horizontal gradients, the effect of different boundary conditions
was also demonstrated in \cite{r:tlh} to apply to a laterally
heated stably stratified slot (LHSSS). The latter
configuration had been studied before only
in the context of the classical
double-diffusion \cite{r:ths}.

The above effect of boundary conditions was also demonstrated in
\cite{r:trp} to result in steady convection when two compensating horizontal 
gradients maintained by different boundary conditions are applied to a
vertical layer of Boussinesq fluid at rest. In the latter configuration,
however, finite-amplitude convective steady flows arise without the
respective linear instability of the conduction state, as in the scenario
first proposed in \cite{r:rd}. This is qualitatively different from the
analogous problem of the classical double-diffusion \cite{r:xqt}, where
the respective linear instability is concomitant of subcritical steady
convection. The finite-amplitude manifestation of convection uncovered
in \cite{r:trp} was also found to take place under a stable vertical
solute stratification. In addition, the scenario in \cite{r:trp}
unveiled the existence of an oscillatory linear instability. It
thus suggested the possibility of a new oscillatory
manifestation of convection due to different
boundary conditions.

Differential diffusion due to unequal perturbation gradients of the
components (differential gradient diffusion \cite{r:wel,r:twh,r:tbc,r:tlh,r:trp})
is not the only process by means of which different boundary
conditions can trigger convection from the state of rest. As reported in
\cite{r:tegs,r:tpla}, finite-amplitude steady convection could arise well
before the onset of the respective linear instability in the viscous version
of the problem considered in \cite{r:wel}. In this case, the disparity
between component stratifications resulting from finite-amplitude
perturbation generates convective flows due to the feedback coming
from nonlinear \mbox{Rayleigh}---\mbox{Benard} convection. Such
finite-amplitude convection was found in \cite{r:tpla}
to coexist with an oscillatory linear instability
in viscous fluid.

The main objective of the present study is to provide a comprehensive
understanding of the manifestation of infinite-slot convection resulting
from different boundary conditions for diverse slot orientations to the
gravity. The central issue in achieving this objective is oscillatory
small-amplitude manifestation of such convection. Resulting from this
work, the understanding of this issue has required uncovery and analysis
of inviscid and viscous transformations of the oscillatory instability
into respective steady one. In addition, it required identification of a new
physical mechanism of oscillatory instability. Such instability arises in the
vertical slot with viscous fluid \cite{r:trp}. The key physical effects underlying
transformation between the inviscid mechanism of horizontal-slot oscillatory
instability in \cite{r:wel} and the new vertical-slot mechanism in viscous
fluid are thus also described in this work. Implications of the provided
interpretations, in particular for the existence of a three-dimensional
oscillatory instability in an inclined slot, are analyzed as well. This
study also completes the understanding of steady convection due to
different boundary conditions when a horizontal slot is
transformed into an inclined and vertical slot.
\section{\label{s:f}The problem formulation and solution procedures}
\subsection{\label{s:fg}The problem and governing equations}
\begin{figure}
\centerline{\psfig{file=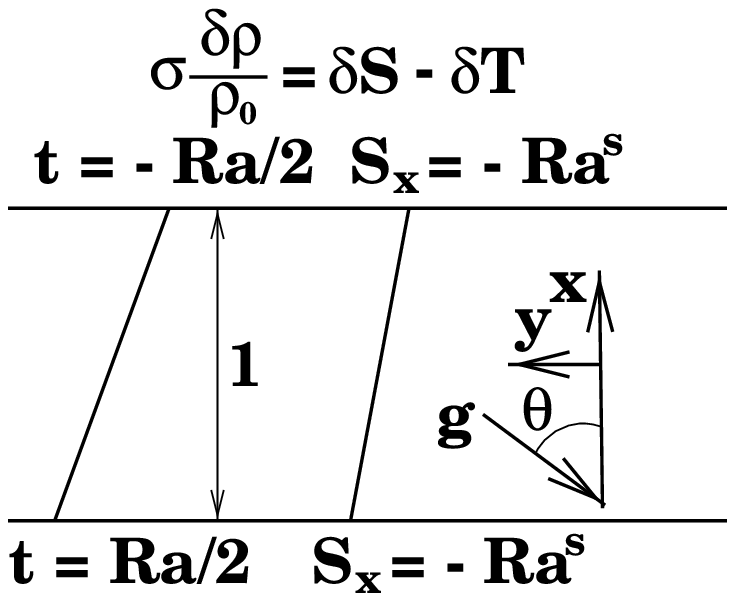,width=8.25cm}}
\vspace*{0.5cm}
\caption{Geometry of the problem. $\delta\rho=\rho-\rho_{0}$ 
is the variation of the (dimensionless) density, $\rho$, due to the variations $\delta S$ and $\delta T$ of
solute concentration $S$ and temperature $T=(T_1+T_2)/2+t$ with respect to their reference values, at which
the density is $\rho_0$; $T_1$ and $T_2$ are the boundary temperatures, $\sigma=gd^3/\kappa\nu$. 
$Pr\equiv\nu/\kappa=6.7$, $Le\equiv\kappa_{T}/\kappa_{S}=1$; $\kappa_{T}$ and $\kappa_{S}$ ( $= \kappa$)
are the component diffusivities. The fluid is of the Boussinesq type.}
\label{f:g}
\end{figure}

Let compensating gradients of the contributions of two components to the density
be maintained by different boundary conditions at the walls of a slot with pure
fluid whose orientation to the gravity is characterized by angle $\theta$. This
is a particular case of the problem illustrated in Fig. \ref{f:g}, where
$\theta$ ( $> 0$ in Fig. \ref{f:g}) is the angle between the direction opposite to the gravity
and that of the across-slot coordinate axis. The component gradients in Fig. \ref{f:g}
are represented by the \mbox{Rayleigh} numbers $Ra=g\alpha\Delta \overline{T}d^3/\kappa\nu$
and $Ra^{s}=-g\beta(\partial\overline{S}/\partial\overline{x})d^4/\kappa\nu\equiv\mu Ra$.
Here, $\overline{x}$ is the (dimensional) across-slot coordinate, $d$ is the width of the
slot, $\Delta\overline{T}$ is the (dimensional) difference between the values of temperature
(standing for the component with fixed-value boundary conditions) at the boundaries with
smaller and larger across-slot coordinates, $\partial \overline{S}/\partial\overline{x}$
is the boundaries-prescribed (dimensional) derivative of solute concentration
(standing for the component with flux boundary conditions), $\alpha$ is the coefficient
of thermal expansion, $\beta$ is the coefficient of the density variation due to the
variation of solute concentration, $g$ is the gravitational acceleration, $\nu$ is
the kinematic viscosity, and $\kappa=\kappa_{T}=\kappa_{S}$ is the diffusivity
of both components. The bar means that the respective variable is
dimensional. Unless explicitly stated otherwise, $Ra>0$
and $Ra^{s}>0$ as well as $\mu=1$ are assumed.

The component diffusivities are set equal to eliminate the effects of the
classical double-diffusion. The effects of different boundary conditions could
thus be analyzed separately from those of unequal diffusivities. Such approach
is analogous to that used in previous studies of conventional double-diffusive
convection. In most such studies, the components with unequal diffusivities
have not been distinguished from each other in terms of boundary conditions.
As indicated in \cite{r:pmsns}, equal diffusivities can in principle be
experimentally modeled with two solutes. The \mbox{Prandtl} number would
then be significantly different from the $Pr=6.7$ specified in the caption
of Fig. \ref{f:g}. This parameter, however, is not expected to have a
qualitative effect on the results and physical interpretations reported
herein. Equal diffusivities could also be interpreted as eddy transport
coefficients, as in \cite{r:l83th04,r:qgdm}. In this case, $Pr=6.7$
is not outside the range of realistic values corresponding
to all diffusion coefficients being
of the eddy type.

The condition that the component
gradients be exactly compensating ($\mu=1$) eliminates
the along-slot motion arising when the slot orientation
differs from horizontal. (This condition has also been adopted in many
previous studies of conventional double-diffusive convection \cite{r:xqt,r:bgm}.)
In the absence of such a background flow, comparison of the results for an inclined
or vertical slot with those for the horizontal slot is substantially
facilitated. The same could also be said about the comparison between
the viscous and inviscid problems, for the along-slot base flows
would have been dissimilar in such problems. In particular, these
comparisons are relevant for understanding the physics of
transformation between oscillatory convection in inviscid
fluid at $\theta=0$ \cite{r:wel} and that in viscous
fluid at $\theta=\pi/2$ \cite{r:trp}. 

The equations describing the two-dimensional
(2D) problem in Fig. \ref{f:g} in which an along-slot
solute stratification is also present can be
written as follows:
\begin{displaymath}
\frac{\partial\zeta}{\partial\tau}+
\frac{\partial\psi}{\partial x}\frac{\partial\zeta}
{\partial y}-\frac{\partial\psi}{\partial y}
\frac{\partial\zeta}{\partial x}=
\frac{1}{Pr}(\frac{\partial t}{\partial x}-
\frac{\partial s}{\partial x})\sin\theta-    
\end{displaymath}
\begin{equation}
\frac{1}{Pr}(\frac{\partial t}{\partial y}-
\frac{\partial s}{\partial y})\cos\theta+       
\frac{\partial^2\zeta}{\partial x^2}+
\frac{\partial^2\zeta}{\partial y^2},       \label{eq:ns1}
\end{equation}
\begin{equation}
\zeta=\frac{\partial^2\psi}{\partial x^2}+
\frac{\partial^2\psi}{\partial y^2},         \label{eq:ns2}
\end{equation}
\begin{equation}
\frac{\partial t}{\partial\tau}+\frac{\partial\psi}{\partial x}\frac{\partial t}
{\partial y}-\frac{\partial\psi}{\partial y}
\frac{\partial t}{\partial x}=
\frac{1}{Pr}(\frac{\partial^2 t}{\partial x^2}+
\frac{\partial^2 t}{\partial y^2}),         \label{eq:dt}
\end{equation}
\begin{equation}
\frac{\partial s}{\partial\tau}+\frac{\partial\psi}{\partial x}
(\frac{\partial s}{\partial y}-Ra_{S})-
\frac{\partial\psi}{\partial y}\frac{\partial s}{\partial x}=
\frac{1}{Pr}(\frac{\partial^2 s}{\partial x^2}+
\frac{\partial^2 s}{\partial y^2}),         \label{eq:ds}
\end{equation}
where the across-slot, $u$, and along-slot, $v$, velocities are 
$$u=-\frac{\partial\psi}{\partial y},\;\;\;\;\;\;\;
v=\frac{\partial\psi}{\partial x},$$
vorticity $$\zeta=\frac{\partial v}{\partial x}-
\frac{\partial u}{\partial y},$$
solute concentration
$$S=-Ra_{S}y+s,$$
temperature
$$T=\frac{T_{1}+T_{2}}{2}+t,$$
$Pr=\nu/\kappa$ is the \mbox{Prandtl} number, $\tau$ is
the time, $x\in(-1/2,1/2)$, $y\in(-\lambda/2,\lambda/2)$,
and $\lambda=\overline{\lambda}/d$ is the specified
along-slot period.

Unless explicitly emphasized otherwise,
the along-slot solute stratification, characterized
by the \mbox{Rayleigh} number $Ra_{S}=g\beta|\partial \overline{S}/
\partial\overline{y}|d^4/\kappa\nu$, is assumed to be zero. The
formulation with $Ra_{S}\neq 0$ is used only for facilitating
the discussion related to results in \cite{r:tlh,r:trp},
where the problems with $Ra_{S}\neq 0$ have been 
addressed.

The above equations were considered along with the
no-slip formulation of wall boundary conditions
\begin{displaymath}
\zeta=\frac{\partial^{2}\psi}{\partial x^{2}},\;\;\;\;\;
\psi=0,\;\;\;\;\;t=\pm\frac{Ra}{2},\;\;\;\;\;
\end{displaymath}
\begin{equation}
\frac{\partial s}{\partial x}=-\mu Ra=-Ra^{s}\;\;
(x=\mp 1/2,\;-\lambda/2<y<\lambda/2) \label{bc:lr}
\end{equation}
and periodic boundary conditions in the along-slot direction
\begin{displaymath}
\xi(x,\lambda/2)=\xi(x,-\lambda/2),\;\;\;\;\;\;
\frac{\partial\xi(x,\lambda/2)}{\partial y}=
\frac{\partial \xi(x,-\lambda/2)}{\partial y}
\end{displaymath}
\begin{displaymath}
(-1/2<x<1/2), \label{bc:pbcxi}
\end{displaymath}
\begin{displaymath}
s(x,\lambda/2)=s(x,-\lambda/2),\;\;\;\;\;\;\;
\frac{\partial s(x,\lambda/2)}{\partial y}=
\frac{\partial s(x,-\lambda/2)}{\partial y}
\end{displaymath}
\begin{equation}
(-1/2<x<0,\;\;0<x<1/2),\;\;\;\;
s(0,\pm\lambda/2)=0,\;\;\;\; \label{bc:pbcs}
\end{equation}
where $\xi$ stands for $\zeta$, $\psi$, and $t$. As indicated in
the caption of Fig. \ref{f:g}, $Pr=6.7$ was used throughout the
present study. As at the middle points of across-slot boundaries
in (\ref{bc:pbcs}), specification of the values of $s$ is needed
to identify the solute concentration scale and select the phase
of a nontrivial flow. Technical difficulties that still
exist with the employed approach will be discussed
in Sec. \ref{s:rsng}.

The steady version of Eqs. (\ref{eq:ns1})---(\ref{eq:ds})
and boundary conditions (\ref{bc:lr}) and (\ref{bc:pbcs}) was
discretized by central finite differences. The \mbox{Euler---Newton} and
\mbox{Keller} \cite{r:kel} arclength continuation algorithms were used
\cite{r:tlh} to trace out bifurcating branches of steady convection using
the Harwell MA32 Fortran routine. In all these computations, the along-slot
period $\lambda=2$ was prescribed. The grid with $33$ nodes in the across-slot
direction was used in the computations [$nx\times\lambda(nx+1)$ with $nx=33$],
as in \cite{r:twh,r:tlh,r:trp,r:tpla}. Temporal behavior of the linearized version
of Eqs. (\ref{eq:ns1})---(\ref{eq:ds}) and boundary conditions (\ref{bc:lr})
and (\ref{bc:pbcs}) was also examined, in particular near the onset
of oscillatory instability of the conduction state. For this purpose, the
implicit method was employed to compute time evolution of the
respective linear system with time step $\delta\tau=0.05$.
\subsection{\label{s:fl}Linear stability calculations}
\subsubsection{\label{s:fl2}Two-dimensional disturbances}
With the state of rest being the background flow for $\mu=1$,
the \mbox{Fourier} mode of a 2D marginally unstable
oscillatory perturbation with angular frequency
$\omega$ and wave number $k$ can be written as 
\begin{equation}
[u'(x),t'(x),s'(x)]^{T}
e^{i(\omega\tau\pm k y)}+cc. \label{e:p}
\end{equation}
Here $[u'(x),t'(x),s'(x)]^{T}$ is the \mbox{Fourier}-mode part depending on
the across-slot coordinate alone, the prime near a flow variable denotes such
part in the perturbation of the variable. Expression (\ref{e:p}) has been
introduced into the linearized Eqs. (\ref{eq:ns1})---(\ref{eq:ds})
rewritten for $Ra_{S}=0$ in terms of across-slot velocity $u$,
relative temperature $t$, and solute concentration $s$.
This leads to:
\begin{displaymath}
(\frac{d^2}{dx^2}-k^2)(\frac{d^2}{dx^2}-k^2-i\omega)\tilde{u}=
\end{displaymath}
\begin{equation}
\mp Ra[ik\frac{d}{dx}(\tilde{t}-\tilde{s})\sin\theta\pm
k^2(\tilde{t}-\tilde{s})\cos\theta], \label{eq:lsu}
\end{equation}
\begin{equation}
(\frac{d^2}{dx^2}-k^2-i\omega Pr)\tilde{t}=
\tilde{u}, \label{eq:lst}
\end{equation}
\begin{equation}
(\frac{d^2}{dx^2}-k^2-i\omega Pr)\tilde{s}=
\tilde{u},             \label{eq:lss}
\end{equation}
where $\tilde{u}=-u'Pr$, $\tilde{t}=t'/Ra$, 
and $\tilde{s}=s'/Ra^{s}$ ($Ra^{s}=Ra$).

The variables $\tilde{u}$, $\tilde{t}$,
and $\tilde{s}$ are subject to the following 
set of boundary conditions:
\begin{equation}
\tilde{u}=\frac{d^{2}\tilde{u}}{dx^{2}}=\tilde{t}=\frac{d\tilde{s}}{dx}=0\;\;\;\; 
(x=\pm 1/2) \label{bc:vs}
\end{equation}
when the boundaries are stress-free. For no-slip
boundaries, the set of boundary conditions
is as follows:
\begin{equation}
\tilde{u}=\frac{d\tilde{u}}{dx}=\tilde{t}=\frac{d\tilde{s}}{dx}=0\;\;\;\; 
(x=\pm 1/2). \label{bc:vn}
\end{equation}

For finding the marginally stable values of $Ra$ and $\omega$, it is sufficient
to consider only positive values of $\omega$ and $k$ in one such traveling wave
as (\ref{e:p}). If $[\tilde{u}(x),\tilde{t}(x),\tilde{s}(x)]^{T}$ is a solution
of the mode $e^{i(\omega\tau+k y)}$ version of Eqs. (\ref{eq:lsu})---(\ref{eq:lss})
at some $Ra_{c}$ and $\omega_{c}$, then the mode $e^{i(\omega\tau-k y)}$ version of
these equations would be satisfied by $\pm[\tilde{u}(-x),\tilde{t}(-x),\tilde{s}(-x)]^{T}$ 
at these same $Ra_{c}$ and $\omega_{c}$. It is thus only the mode $e^{i(\omega\tau+k y)}$
version of linear stability equations that is hereafter implied when
such equations are referred to.

For examination of the linear stability of the conduction state
in inviscid fluid,
\begin{equation}
i\omega(\frac{d^2}{dx^2}-k^2)\tilde{u}=
Ra[ik\frac{d}{dx}(\tilde{t}-\tilde{s})\sin\theta+
k^2(\tilde{t}-\tilde{s})\cos\theta] \label{eq:lsui}
\end{equation}
was used along with
\begin{equation}
(\frac{d^2}{dx^2}-k^2-i\omega)\tilde{t}=
\tilde{u} \label{eq:lsti}
\end{equation}
and 
\begin{equation}
(\frac{d^2}{dx^2}-k^2-i\omega)\tilde{s}=
\tilde{u},             \label{eq:lssi}
\end{equation}
where the \mbox{Rayleigh} numbers are defined as
$Ra=g\alpha\Delta \overline{T}d^3/\kappa^{2}$ and $Ra^{s}=
-g\beta(\partial\overline{S}/\partial\overline{x})d^4/\kappa^{2}=Ra$.
Here $\omega$ is nondimensionalized with $\kappa/d^2$
as opposed to $\nu/d^2$ in Eqs.~(\ref{eq:lsu})---(\ref{eq:lss}).
Although the definitions of $Ra$, $Ra^s$, and $\omega$ for inviscid
fluid are different from the respective definitions for viscous fluid,
the same notations are used for these parameters. It is implied below
that the definitions of $Ra$ and $Ra^{s}$ as well as the time
nondimensionalization scale (in $\omega$) correspond
to the type of fluid in question.

The boundary conditions for inviscid fluid are
\begin{equation}
\tilde{u}=\tilde{t}=\frac{d\tilde{s}}{dx}=0\;\;\;\; 
(x=\pm 1/2). \label{bc:i}
\end{equation}

The inviscid problem at $\theta=0$ was also examined for the $Ra$
and $Ra^{s}$ being independent of each other. That is, equation
\begin{equation}
i\omega(\frac{d^2}{dx^2}-k^2)\tilde{u}=
k^2(Ra\tilde{t}-Ra^{s}\tilde{s}) \label{eq:lsuii}
\end{equation}
was considered along with Eqs.~(\ref{eq:lsti}) and
(\ref{eq:lssi}) and boundary conditions (\ref{bc:i})
for specified values of $Ra^{s}$.

Eqs. (\ref{eq:lst}) and (\ref{eq:lss}) are first solved for $\tilde{t}-\tilde{s}$,
upon which the resulting general solution is introduced into Eq.~(\ref{eq:lsu}).
The latter equation is then solved for $\tilde{u}$, and the general
expressions for $\tilde{t}$ and $\tilde{s}$ are thus obtained from
Eqs.~(\ref{eq:lst}) and (\ref{eq:lss}). For a fixed $k$, $Ra_{c}$
and $\omega_{c}$ are then found by searching in the $Ra$---$\omega$
domain for the smallest $Ra$ at which the complex matrix resulting
from the application of boundary conditions (\ref{bc:vs}) or
(\ref{bc:vn}) to the obtained general solution is singular.
The same procedure was also applied to the general solution
of either Eqs. (\ref{eq:lsui})---(\ref{eq:lssi}) or Eqs.
(\ref{eq:lsti}), (\ref{eq:lssi}), and (\ref{eq:lsuii})
along with boundary conditions (\ref{bc:i}). NAG
Fortran routines were employed for this purpose. 

Once $Ra_{c}(k_0)$ and $\omega_{c}(k_0)$ have been found for
a given $k_{0}$, the corresponding values of these parameters, 
$Ra_{c}(k)$ and $\omega_{c}(k)$, at a nearby $k=k_{0}+\delta k$
can be computed by the \mbox{Euler}---\mbox{Newton} continuation
method. This method was applied to the solution of equation
\begin{equation}
F[Ra(k),\omega(k),k]=0, \label{eq:sm}
\end{equation}
where $F[Ra(k),\omega(k),k]$ stands for the (complex) determinant of the
matrix resulting from the application of boundary conditions (\ref{bc:vs}),
(\ref{bc:vn}), or (\ref{bc:i}) to the general solution of the respective set
of differential equations. As mentioned above, such determinants were computed with the
use of standard Fortran routines. Numerical differentiation was thus employed
to compute the Jacobian of $\{Re[F(Ra,\omega,k)],Im[F(Ra,\omega,k)]\}^{T}$ (with
respect to $Ra$ and $\omega$) and $\partial{F(Ra,\omega,k)}/\partial{k}$. 

One also needs
to ensure that no disconnected branches of $[Ra(k),\omega(k)]^{T}$
satisfying the real and imaginary parts of Eq. (\ref{eq:sm}) arise
for the values of $Ra$ that are smaller than those computed by the
continuation procedure. For this purpose, $Ra_{c}(k)$ and $\omega_{c}(k)$
were independently obtained for several values of $k$ by the search
in the $Ra$---$\omega$ domain outlined above. In this case, however,
the values of $k$ could be relatively scarcely spaced within the
considered wave number interval. The results of such searches
were found to be entirely consistent with those obtained
by the \mbox{Euler}---\mbox{Newton} continuation.

For viscous fluid, the linear stability to steady disturbances
was also considered, as $\theta\in(\pi/2,\pi)$. For this purpose,
$\omega=0$ was set in Eqs. (\ref{eq:lsu})---(\ref{eq:lss})
and the general solution of (resulting) equations
\begin{equation}
(\frac{d^2}{dx^2}-k^2)^{2}\tilde{u}=
-Ra[ik\frac{d}{dx}(\tilde{t}-\tilde{s})\sin\theta+
k^2(\tilde{t}-\tilde{s})\cos\theta], \label{eq:lssu}
\end{equation}
\begin{equation}
(\frac{d^2}{dx^2}-k^2)\tilde{t}=
\tilde{u}, \label{eq:lsst}
\end{equation}
and
\begin{equation}
(\frac{d^2}{dx^2}-k^2)\tilde{s}=
\tilde{u}  \label{eq:lsss}
\end{equation}
was obtained analytically. Either boundary conditions
(\ref{bc:vs}) or (\ref{bc:vn}) were then applied to this
general solution and the smallest $Ra$, $Ra_c(k)$, at which
the resulting matrix becomes singular were searched for at
different $k$.
\subsubsection{\label{s:fl3}Three-dimensional disturbances}
For $\theta\neq 0,\pi$, a three-dimensional (3D) structure of
perturbation can generally become relevant in the linear stability
analysis. This study is mainly focused on 2D disturbances, for which
understanding of the basic physical effects of different boundary
conditions can be most conveniently obtained. Unless explicitly
stated otherwise, therefore, it is the 2D linear instability
that is implied below. However, 3D disturbances were also
considered: in some cases, their analysis was found
to be useful for understanding the 2D results. 

With the same nondimensionalization as in $\tilde{u}$, $\tilde{t}$,
and $\tilde{s}$, the \mbox{Fourier} mode of a 3D disturbance
is described as
\begin{equation}
[\tilde{u}(x),\tilde{t}(x),\tilde{s}(x)]^{T} 
e^{i(\omega\tau+k_{y}y+k_{z}z)}+cc, \label{e:p3}
\end{equation}
where $k_{y}$ and $k_{z}$ are the $y$ and $z$ components, respectively,
of the full wave number $k=(k_{y}^2+k_{z}^2)^{1/2}$. (The $z$-axis is
orthogonal to the $x$---$y$ plane in Fig. \ref{f:g} and is directed
towards the reader.) Expression (\ref{e:p3}) was then introduced
into the appropriately nondimensionalized, linearized 3D
governing equations for viscous and inviscid fluid.

With the definition of wave number $k$ just generalized for 3D
disturbances, the marginal linear stability in viscous fluid
is described by the equations formally identical to Eqs.
(\ref{eq:lst}) and (\ref{eq:lss}) along with
\begin{displaymath}
(\frac{d^2}{dx^2}-k^2)(\frac{d^2}{dx^2}-k^2-i\omega)\tilde{u}=
-Ra[ik_{y}\frac{d}{dx}(\tilde{t}-\tilde{s})\sin\theta+
\end{displaymath}
\begin{equation}
k^2(\tilde{t}-\tilde{s})\cos\theta]. \label{eq:lsu3}
\end{equation}
Likewise,
the 3D marginal stability in inviscid fluid is described
by the equations formally identical to Eqs. (\ref{eq:lsti})
and (\ref{eq:lssi}) along with
\begin{equation}
i\omega(\frac{d^2}{dx^2}-k^2)\tilde{u}=
Ra[ik_{y}\frac{d}{dx}(\tilde{t}-\tilde{s})\sin\theta+
k^2(\tilde{t}-\tilde{s})\cos\theta]. \label{eq:lsui3}
\end{equation}
For fixed values of $k_{z}$, $Ra_{c}(k_{y})$ and $\omega_{c}(k_{y})$
were obtained with the same procedure as described above
for 2D disturbances.
\section{\label{s:ro}Oscillatory convection}
\subsection{\label{s:rog}General}
For interpretation of the findings on oscillatory instability,
only a reflectionally symmetric oscillatory perturbation is considered
below. In a horizontal slot, such perturbations are often referred to as
standing waves. They are then characterized by convection cells changing
their sense of rotation periodically in time. As suggested by \cite{r:ckgs},
both reflectionally and translationally symmetric oscillatory perturbations
are expected to arise from the \mbox{Hopf} bifurcation in a system whose
background state possesses both these symmetries. Translationally symmetric
oscillatory perturbations are often referred to as traveling waves. However,
it is sufficient to understand the physical mechanism of such oscillatory
instability in terms of only one of the types of perturbation just
mentioned. Amplitude growth of the oscillatory perturbation of
the other type can then be viewed merely as a mathematical
consequence of the respective results in \cite{r:ckgs}.
\subsection{\label{s:roi}Inviscid fluid}
\subsubsection{\label{s:roih}Effect of the across-slot gravity.}
\begin{figure}
%\vspace*{1cm}
\centerline{\psfig{file=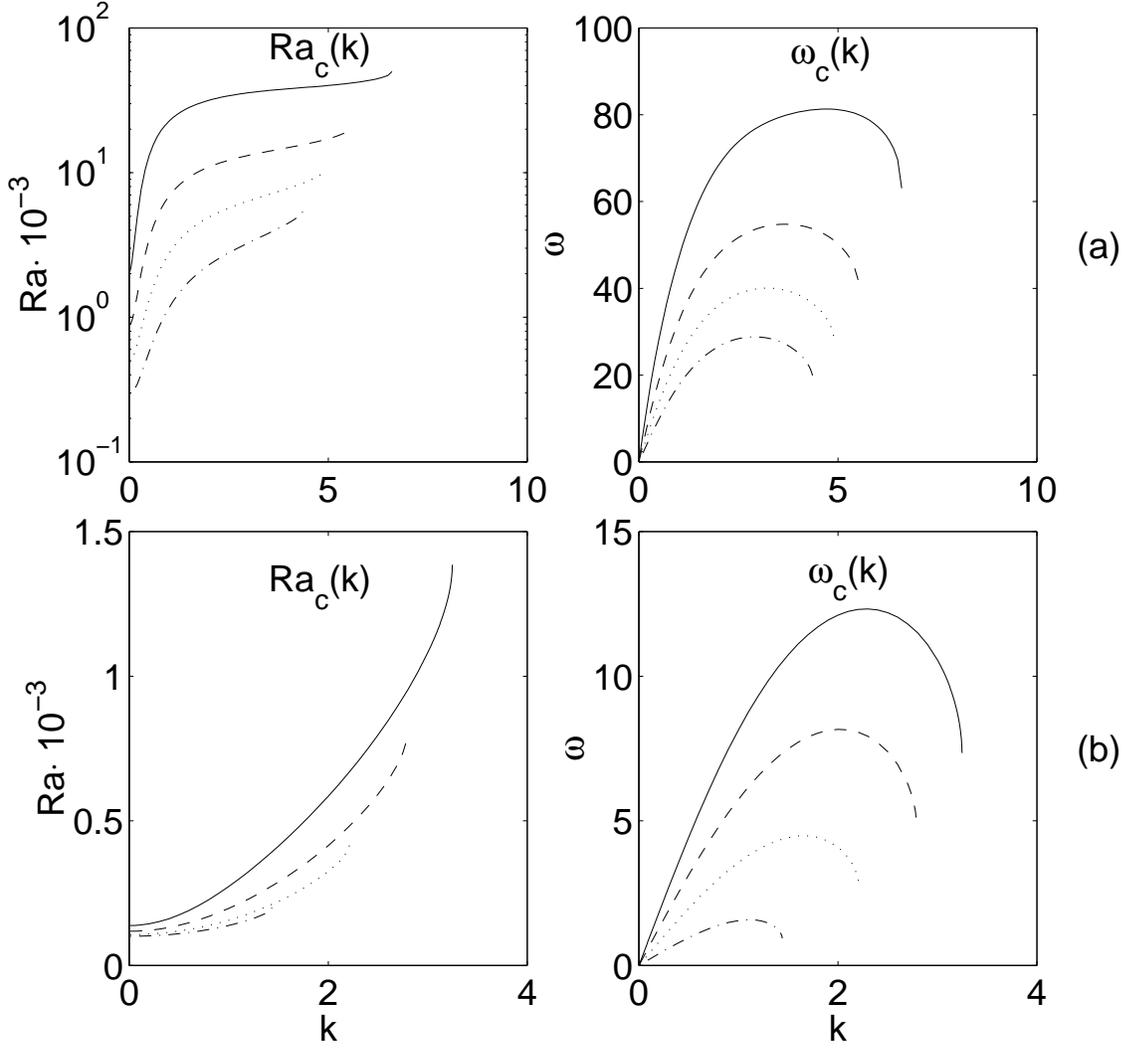,width=15cm}}
\vspace*{0.5cm}
\caption{$\theta=0$. Inviscid fluid. Curves of the marginal linear stability
to 2D oscillatory disturbances for independently prescribed $Ra^{s}$, $Ra_{c}(k)$
and $\omega_{c}(k)$; $Le=1$. (a) the solid lines: $Ra^{s}=50000$, the dashed lines:
$Ra^{s}=20000$, the dotted lines: $Ra^{s}=10000$, the dash-dot lines: $Ra^{s}=5000$;
(b) the solid lines: $Ra^{s}=1000$, the dashed lines: $Ra^{s}=500$,
the dotted lines: $Ra^{s}=200$, the dash-dot lines: $Ra^{s}=50$.}
\label{f:mscoirss}
\end{figure}

Let us first consider the problem in a horizontal slot addressed in \cite{r:wel}
($\theta=0$), whose linear stability is described by Eqs. (\ref{eq:lsti}),
(\ref{eq:lssi}), and (\ref{eq:lsuii}) and boundary conditions (\ref{bc:i}).
The marginal-stability curves, $Ra_{c}(k)$ and $\omega_{c}(k)$, are illustrated
in Fig. \ref{f:mscoirss}. They exhibit two basic features. The first is that
the most unstable wave number is zero for any $Ra^{s}$. This wave number
is also characterized by zero oscillation frequency. The second feature
is the existence of a minimal horizontal wavelength, decreasing
with the increase of $Ra^{s}$, below which the
instability does not arise. 

Reinterpreting \cite{r:wel}, rotation of a small-amplitude perturbation
cell generates a potential energy of component perturbation stratifications
in the end of a rotation cycle. Due to differential gradient diffusion, this
potential energy is utilized by the perturbation cell in the beginning of the
cycle of cell rotation in the opposite sense. The maximal amount of such energy
depends on the time available for a fluid element of the cell to change
its vertical coordinate. This time is specified by the horizontal scale
of the instability. For this reason, $Ra_{c}(k)$ decreases with the
increase of the horizontal wavelength and assumes its minimal
value for a given $Ra^{s}$ as $k\rightarrow 0$ (i.e., as
wavelength $\lambda\rightarrow\infty$). 

Below a certain critical
wavelength, the potential energy utilized
by a perturbation cell becomes insufficient for bringing about growth
of the perturbation amplitude. The increase of $Ra^{s}$ and associated
growth of $Ra_{c}$ enhance the disparity between the respective
gradients in the perturbed state. This makes the energy transfer
to the perturbation cells more intensive. The minimal unstable
wavelength thus decreases as $Ra^{s}$ grows
(Fig. \ref{f:mscoirss}).

Efficiency of the utilization of the potential energy by a perturbation cell 
depends on the frequency with which the marginally unstable cells change their sense of
rotation. This frequency thus has to be such that the time for vertical diffusion naturally
specified by the instability wavelength be resonantly matched. Since such time grows with
the wavelength increase, $\omega_{c}(k)\rightarrow 0$ as $k\rightarrow 0$ and $\omega_{c}(k)$
grows with the increase of $k$ from $0$. As the wave number is further increased, the
wavelength time for the manifestation of differential diffusion eventually becomes
insufficient for the cell oscillation amplitude to grow. As a consequence, 
$\omega_{c}(k)$ begins to decrease when certain values of $k$ are
exceeded (Fig. \ref{f:mscoirss}), to afford
more time for the diffusion.

The decrease of $\omega_{c}(k)$, however, leads to an inconsistency between
the diffusion time afforded by the perturbation frequency and that specified by
the instability wavelength. This results in the efficiency of utilization of
the component potential energy by a perturbation cell being reduced. The
additional energy that can actually be utilized due to such frequency
decrease is thus expected to be limited. As the wave number
exceeds a critical value, therefore, the instability
fails to develop.

The fact that the most unstable wave number is zero makes the determination of exact values
of $Ra_{c}(0)$ and group velocity $\omega^{c}_{k}(0)\equiv\partial{\omega_{c}(0)}/\partial{k}$
relevant. Also applicable to Eqs. (\ref{eq:lsui})---(\ref{eq:lssi}) for $\theta>0$,
the long-wavelength expansion used for this purpose is as follows:
\begin{displaymath}
Ra_{c}(k)=Ra_{0}+k^2Ra_{2}+...,
\end{displaymath}
\begin{displaymath}
\omega_{c}(k)=k(\omega_{0}+k^2\omega_{2}+...),
\end{displaymath}
\begin{displaymath}
\tilde{u}(k)=k(\tilde{u}_0+k\tilde{u}_1+k^2\tilde{u}_2+...),
\end{displaymath}
\begin{displaymath}
\tilde{t}(k)=\tilde{t}_0+k\tilde{t}_1+k^2\tilde{t}_2+...,
\end{displaymath}
\begin{equation}
\tilde{s}(k)=\tilde{s}_0+k\tilde{s}_1+k^2\tilde{s}_2+.... \label{eq:lwe}
\end{equation}

Upon introduction of (\ref{eq:lwe}) into Eqs. (\ref{eq:lsti}),(\ref{eq:lssi}), and
(\ref{eq:lsuii}), and use of boundary conditions (\ref{bc:i}), $\tilde{t}_{0}$,
$\tilde{s}_{0}$, and $\tilde{u}_{0}$ are first obtained. The $k^{1}$ order
of Eq.  (\ref{eq:lssi}) and the flux-free boundary conditions
for $\tilde{s}_{1}$ from (\ref{bc:i}) then yield
\begin{equation}
\omega_{k}^c(0)\equiv\frac{\partial\omega_c(0)}{\partial k}=\omega_0=\sqrt{Ra^{s}/12}. \label{eq:omc}
\end{equation}
Having derived the expressions for $\tilde{s}_{1}$ and $\tilde{t}_{1}$ and on introduction of these
into the $k^{1}$ order of Eq. (\ref{eq:lsuii}), one obtains the expression for $\tilde{u}_{1}$.
With such $\tilde{u}_{1}$, the $k^{2}$ order of Eq. (\ref{eq:lssi}) and the flux-free
boundary conditions for $\tilde{s}_{2}$ from (\ref{bc:i}) thus yield

\begin{equation}
Ra_c(0)=Ra_0=(2Ra^s+5040)/51. \label{eq:rac}
\end{equation}
The numerical data underlying the marginal-stability curves in
Fig. \ref{f:mscoirss} were found to accurately coincide
with Eqs. (\ref{eq:omc}) and (\ref{eq:rac}).
For the case of two compensating gradients this work is aimed at, $Ra_{c}(0)=Ra^{s}$, 
\begin{equation}
Ra_c(0)=Ra_0=12\cdot 60/7,\;\;\; \omega_{k}^{c}(0)=\omega_{0}=\sqrt{60/7}. \label{eq:roc}
\end{equation}
\subsubsection{\label{s:roiv}Effect of the along-slot gravity.}
\begin{figure}
%\vspace*{1cm}
\centerline{\psfig{file=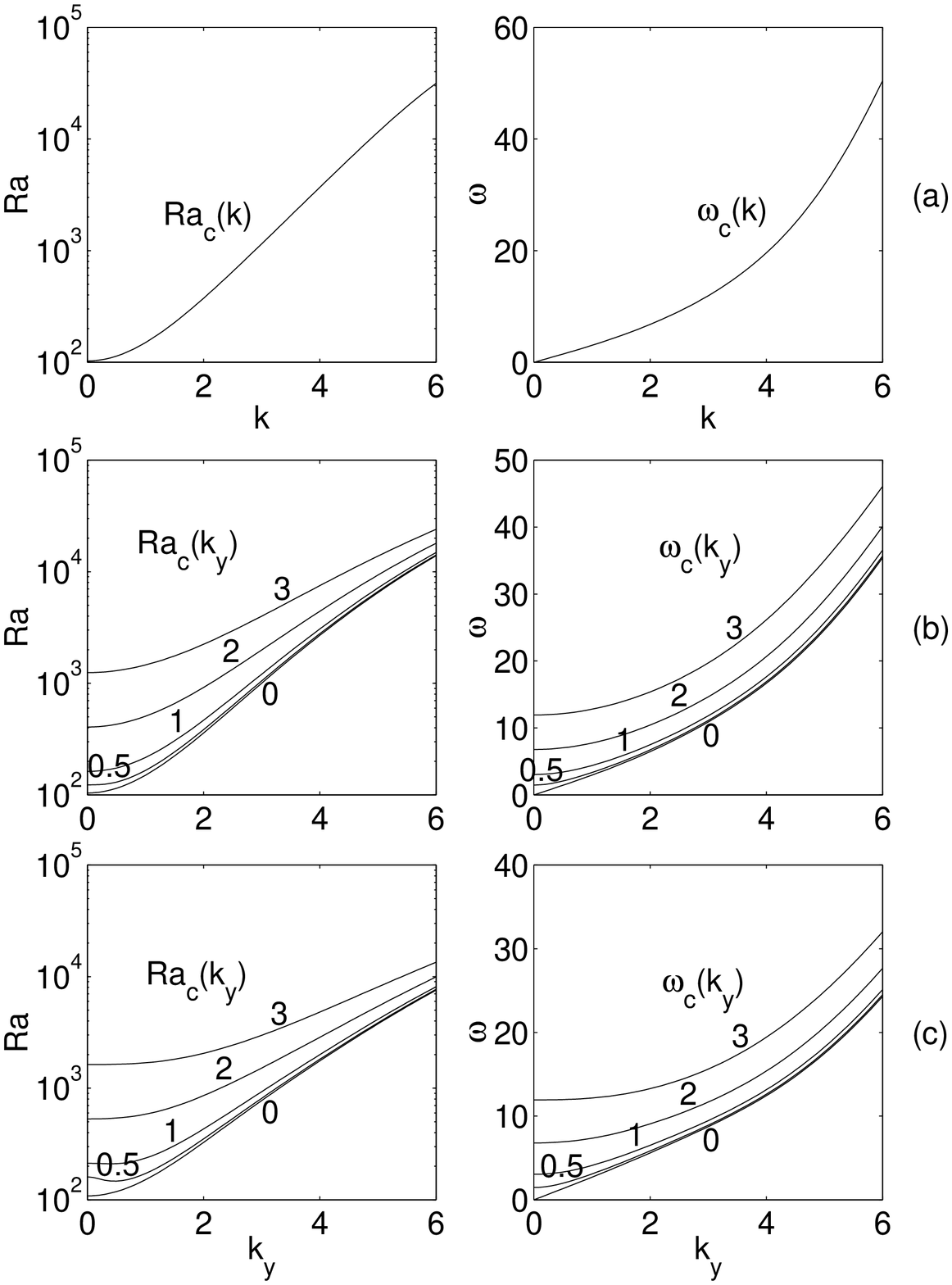,width=15cm}}
%\vspace*{0.5cm}
\caption{Inviscid fluid. Curves of the marginal linear stability to
2D ($k_{z}=0$) oscillatory disturbances, $Ra_{c}(k)$ and $\omega_{c}(k)$,
and to such 3D ($k_{z}\geq 0$) disturbances with different values of
$k_{z}$, $Ra_{c}(k_{y})$ and $\omega_{c}(k_{y})$; $\mu=1$, $Le=1$.
The numbers in (b)---(e), where the disturbances with $k_{z}>0$
were also considered, give the values of $k_{z}$ corresponding
to their nearest curve they do not intersect. (a) $\theta=0$;
(b) $\theta=\pi/8$; (c) $\theta=\pi/4$; (d) $\theta=3\pi/8$;
(e) $\theta=0.95\pi/2$; (f) the dotted lines: $\theta=0.99\pi/2$,
the dashed lines: $\theta=0.995\pi/2$, the solid lines: 
$\theta=0.999\pi/2$.}
\label{f:mscoi}
\end{figure}

For the horizontal and inclined slots with two
compensating across-slot gradients, $Ra_{c}(k)$ and $\omega_{c}(k)$,
obtained from Eqs. (\ref{eq:lsui})---(\ref{eq:lssi}) and boundary conditions
(\ref{bc:i}), are shown in Fig. \ref{f:mscoi}. (In this Sec. \ref{s:roiv},
$k_{z}=0$ and thus also $k=k_{y}$ are implied.) As in the horizontal slot
[Figs. \ref{f:mscoirss} and \ref{f:mscoi}(a)], $k=0$ remains the most
unstable wave number up to the orientation of the slot being nearly vertical
[Figs. \ref{f:mscoi}(b)---\ref{f:mscoi}(e)]. With increasing $\theta$, however,
$Ra_{c}(k)$ and $\omega_{c}(k)$ mostly decrease and qualitative changes arise in
the shape of these curves [Figs. \ref{f:mscoi}(e) and \ref{f:mscoi}(f)]. This
suggests the emergence of an additional factor destabilizing the system.

As suggested by \cite{r:twh,r:tbc}, if $\theta=0$ a
viscous fluid in the region of $Ra<0$ is expected to develop a
steady instability whose mechanism is conceptually analogous to the
finger instability in the classical double-diffusion \cite{r:stnf}. 
Such mechanism is the only cause of linear steady instability for
$0<\theta<\pi/2$ as well (see Sec. \ref{s:rslv} below).
This steady instability would thus have finite negative values of the
marginally unstable \mbox{Rayleigh} number for viscous fluid \cite{r:tbc}
(see also Sec. \ref{s:rslh}). Eqs. (\ref{eq:lssu})---(\ref{eq:lsss})
then suggest that in the inviscid fluid ($Pr\rightarrow 0$), $Ra_{c}(k)=0$
for the respective wave number interval. $Ra<0$ would thus be the unstable
region, with the instability mechanism being similar to that discussed
in \cite{r:twh,r:tbc}. $Ra>0$ is therefore the region of
stability to the effect of the across-slot gravity
component on steady disturbances.

As the along-slot gravity component arises with $\theta$ growing
from $0$, however, another mechanism of differential gradient diffusion
becomes increasingly relevant. This mechanism favors a monotonic growth of
disturbances. For viscous fluid, it has been discussed in \cite{r:tlh,r:trp}.
In the absence of dissipation (throughout this work, this term implies only
viscous dissipation), such mechanism is reasonably expected to be effective
for small-amplitude disturbances at any $Ra\neq 0$. (The results in Fig. \ref{f:mscoi}
discussed just below confirm this assumption.) So long as $\theta<\pi/2$, however,
it should not give rise to steady linear instability for $Ra>0$. The differential
diffusion due to the across-slot gravity component opposes amplitude growth of
the perturbation cells whose sense of rotation does not
change (Sec. \ref{s:roih}).

For $0<\theta<\pi/2$, part of the rotation energy of a perturbation cell
thus comes from the energy directly contributed via the along-slot gravity
component in the current cycle of rotation \cite{r:tlh,r:trp}. In the end
of a cell rotation cycle, the whole rotation energy is transformed into the
potential energy of the perturbation stratification due to the across-slot
gravity component. This potential energy is released in the next rotation
cycle. Depending on the relative roles of the across-slot and
along-slot gravity components, $\omega_{c}(k)$ thus
decrease with increasing $\theta$. 

For $Ra\cos\theta$ being between $0$ and the $Ra_{c}(k)$ from Fig.
\ref{f:mscoi}(a), the growth of oscillatory perturbations is largely
due to the monotonic contribution of the along-slot gravity component, 
effective at any $Ra\neq 0$. The relative role of the across-slot gravity
component in such growth on a given wavelength is specified by the
respective effectiveness of transformation of the energy of cell
rotation into the potential energy of the across-slot perturbation
stratification. (Such process maintains the oscillatory nature
of the instability.) This effectiveness can be judged based
on the behavior of $Ra_{c}(k)$ in Fig. \ref{f:mscoi}(a). 

Compared to the vicinity of the infinite wavelength, finite
wavelengths are less effective in transforming the energy of cell rotation
into the above potential energy. This
makes the relative role of the along-slot gravity component on such finite scales
more pronounced than for the longer wavelengths. [As $\theta$ grows from $0$ in
Figs. \ref{f:mscoi}(a)---\ref{f:mscoi}(d), in particular, $Ra_{c}(k)$ decrease
for $k>\sim 1.1$ and, due to the decrease of the across-slot gravity component,
increase slightly for $k<\sim 1$.] Near $\theta=\pi/2$, a set of finite
wavelengths thus becomes more unstable than the vicinity
of the infinite wavelength [Fig. \ref{f:mscoi}(f)]. 

With the across-slot gravity component vanishing as
$\theta\rightarrow\pi/2$ [Figs. \ref{f:mscoi}(e) and \ref{f:mscoi}(f)], however,
both $Ra_{c}(k)$ and $\omega_{c}(k)$ decrease for all $k$. For $\theta>\pi/2$,
therefore, the slot with inviscid fluid is expected to be unstable to a continuum
of the wavelengths of steady disturbances for any $Ra>0$, and it ought to be
unstable for any $Ra\neq 0$ at $\theta=\pi/2$. As mentioned above, this
would be consistent with the interpretations in \cite{r:tlh,r:trp} if
the absence of dissipation is allowed for. The oscillatory and steady
instabilities due to the respective effects of differential
gradient diffusion in inviscid fluid are thus continuously
transformed into each other around $\theta=\pi/2$.

Upon introduction of (\ref{eq:lwe}) into
Eqs. (\ref{eq:lsui})---(\ref{eq:lssi}), the $k^{1}$ order of
these equations and boundary conditions (\ref{bc:i}) yields
\begin{equation}
Ra_{0}^{2}\sin^{2}\theta+720Ra_{0}\cos\theta-8640\omega_{0}^{2}=0, \label{eq:roci}
\end{equation}
which is consistent with (\ref{eq:omc}) for $\theta=0$. The values
of $Ra_{c}(0)=Ra_{0}$ and $\omega^{c}_{k}(0)=\omega_{0}$ estimated
from the numerical data underlying the (2D) marginal-stability
curves in Fig. \ref{f:mscoi} were found to be consistent
with Eq. (\ref{eq:roci}) to a fairly high  accuracy.
\subsection{\label{s:rov}Viscous fluid}
\subsubsection{\label{s:rovh}Effect of the across-slot gravity.}
The curves of 2D marginal stability for viscous fluid under stress-free
and no-slip boundary conditions are illustrated in Figs. \ref{f:mscovs} and
\ref{f:mscovn}. They were obtained from Eqs. (\ref{eq:lsu})---(\ref{eq:lss})
and boundary conditions (\ref{bc:vs}) and (\ref{bc:vn}), respectively. (In
this Sec. \ref{s:rovh}, $k_{z}=0$ and thus also $k=k_{y}$ are implied.)
The most prominent feature of the 2D curves in Figs. \ref{f:mscovs}
and \ref{f:mscovn}, compared to the above results for inviscid
fluid, is that $k=0$ is not the most unstable wave number
and $\omega_{c}(k)>0$ as $k\rightarrow 0$.

In the context of the effect of the across-slot gravity component, the obtained
stability of the infinite wavelength may at first sight appear to be counterintuitive.
The overall dissipation of perturbation motion is minimized when the wavelength becomes
infinite. Since the effect of differential gradient diffusion is maximized for
$\lambda\rightarrow\infty$, one may think that the infinite wavelength has to
remain most unstable in viscous fluid as well. In particular, $k=0$ is most
unstable to steady disturbances in the viscous fluid
problem identified by $\theta=\pi$ and $Ra>0$
\cite{r:twh,r:tbc,r:nllm}.

The relative stabilization of the long wavelengths [seen in Figs. \ref{f:mscovs}(a)
and \ref{f:mscovn}(a), for example] with respect to the action of the across-slot
gravity component is associated with a special feature of the effect of dissipation.
This feature arises when the instability is oscillatory. Before describing such
feature, however, it is worth illustrating details of the oscillatory
instability mechanism in the horizontal slot.

Fig. \ref{f:perh}(a) represents the stage when the previously accumulated potential
energy of perturbation stratification is released after the sense of cell rotation
has changed. At this stage, cell motion is favored and opposed by the horizontal
solute and temperature perturbation gradients, respectively. Due to differential
(across-slot) gradient diffusion, the temperature opposition to convective
motion in Fig. \ref{f:perh}(a) is overcome. Such motion thus intensifies
[see $\psi$ in Figs. \ref{f:perh}(a) and \ref{f:perh}(b)].

At the current stage of the oscillation period,
the release of the accumulated potential energy is effected via a
mechanism similar to that discussed in \cite{r:twh,r:tbc}, where an
inversely stratified problem was considered. This is particularly
apparent as one compares the present Fig. \ref{f:perh}(a) with
Fig. 2(a) in \cite{r:tbc}. In the present problem, however, the
intensification of convective motion [$\psi$ in Figs. \ref{f:perh}(a) and
\ref{f:perh}(b)] is inevitably accompanied by changing distributions of
the component perturbation isolines [$s$ and $t$ in Fig. \ref{f:perh}(b)].
With the lower background values of the components at the upper wall, the
clockwise (counterclockwise) rotation gives rise to the relatively smaller
respective perturbations in the right (left) part of a rotating cell.
The component perturbation isolines are thus gradually reset so
that cell motion be favored and opposed by the
temperature and solute perturbation stratifications,
respectively [Fig. \ref{f:perh}(c)].

Due to the interaction of a rotating perturbation
cell with flux boundary conditions, the solute perturbation
isolines become more and more vertically oriented, in contrast
to the perturbation isotherms [$s$ and $t$ in Figs. \ref{f:perh}(d)
and \ref{f:perh}(e)]. The resulting differential gradient diffusion acts
against the sense of rotation of a convective cell. This damps cellular
motion [Figs. \ref{f:perh}(c)---\ref{f:perh}(e)]. The energy of cell
rotation is thus transformed into the potential energy of the component
perturbation stratifications. Such potential energy is released
[Figs. \ref{f:perh}(f) and \ref{f:perh}(g)] after
the sense of cell rotation has changed.

The events in Figs. \ref{f:perh}(f)---\ref{f:perh}(j) are thus
qualitatively identical (up to the inverse perturbation sign) to
those in Figs. \ref{f:perh}(a)---\ref{f:perh}(e). The streamline
pattern in Fig. \ref{f:perh}(j) is different from that 
in Fig. \ref{f:perh}(e). This is associated with the
stages of the change in the sense of cell rotation being slightly
different in these figures. Fig. \ref{f:perh}(j) thus shows that
the change of rotation sense is initiated via the formation of
small cells of the opposite sense of rotation near the
boundaries, where differential gradient
diffusion is most effective.

The oscillatory instability thus arises when the potential
energy generated during the rotation of a perturbation cell
in one sense is sufficient to increase the cell amplitude while
the energy is released. The release of potential energy is effected via
differential gradient diffusion during the cell rotation in the opposite
sense. Part of such energy, however, is inevitably spent on dissipation of
along-slot cell motion. This part is the greater the larger
the wavelength is. Efficiency of the (oscillatory instability) feedback
is thus reduced by the dissipation. [Such feedback forms between the
perturbation gradient disparity in one cycle of cell rotation and
the amplitude of rotation (giving rise to such a disparity) in
the next cycle.] The feedback efficiency, therefore, also
tends to zero when the wavelength becomes infinite.

Although increase of the instability wavelength decreases the
overall dissipation, it enhances the role of the unchanged part of
the dissipation (i.e., of the dissipation of along-slot cell motion) in
the oscillatory instability feedback. For sufficiently long wavelengths,
the latter factor dominates the overall effect of dissipation. To avoid the
zero feedback efficiency when $\omega_{c}\rightarrow 0$, $\omega_{c}(k)$ thus
remains greater than zero as $k\rightarrow 0$. In the long-wavelength limit,
therefore, an increasing portion of fluid particles undergoing a cycle of
differential gradient diffusion fail to acquire a sufficient potential
energy for their perturbation amplitude to grow in the next cycle. In
the context of the effect of the across-slot gravity component, this
explains the relative stabilization of long (2D) wavelengths in Figs.
\ref{f:mscovs}(a)---\ref{f:mscovs}(d) and \ref{f:mscovn}(a)---\ref{f:mscovn}(d)
compared to the respective inviscid-fluid problems
[Figs. \ref{f:mscoi}(a)---\ref{f:mscoi}(d)].
\subsubsection{\label{s:rov3}3D effects.}
The above interpretation of the effect of viscous forces on the
oscillatory instability suggests that three-dimensionality of the
disturbances might be relevant in an inclined slot with viscous fluid.
Let the 2D perturbation wavelength, $\lambda_{y}$, be large enough for
the effect of dissipation on efficiency of the instability feedback to
dominate the overall effect of dissipation. Such efficiency increases as
the orientation of the axis of rotation of a convective cell changes and
the total wavelength decreases with the emergence of a $z$ component of the
wave number: $\lambda=2\pi/(k^{2}_{y}+k^{2}_{z})^{1/2}<\lambda_{y}=2\pi/k_{y}$.
The critical \mbox{Rayleigh} number could thus decrease
compared to the 2D problem.

The horizontal-slot linear stability problem depends
only on the wave number modulus. This implies that
\begin{equation}
Ra_{c}(k_{y},k_{z})=Ra_{c}[(k^{2}_{y}+k^{2}_{z})^{1/2},0].
\label{eq:kmod}
\end{equation}
Let $k_{y}$( $> 0$) be fixed, with respect to the selected coordinate system,
in the interval where $Ra_{c}(k_{y})_{|k_{z}=0}$ decreases with increasing
$k_{y}$. Then $Ra_{c}(k_{y})_{|k_{z}>0}<Ra_{c}(k_{y})_{|k_{z}=0}$ for some
$k_{z}>0$. That $Ra_{c}(k_{y})_{|k_{z}>0}>Ra_{c}(k_{y})_{|k_{z}=0}$ for
all positive $k_{y}$ and $k_{z}$ in inviscid fluid also follows
from the $Ra_{c}(k_{y})$ in Fig. \ref{f:mscoi}(a) only
growing with increasing $k_{y}$.

When $\theta>0$, the wave number orientation
becomes physically meaningful. For small $\theta>0$, 
however, $Ra_{c}(k_{y})_{|k_{z}\geq0}$ are still largely determined
by the linear stability equations for $\theta=0$. These describe the
zero-order perturbation expansion in $\theta$ of such equations for
$\theta>0$. When $\theta$ is small, therefore, most unstable
disturbances with small $k_{y}$ are still expected to
be of 3D and 2D nature in viscous and inviscid
fluid, respectively. 

Such a three-dimensionality of instability hinges on the existence of an interval
of growing $k$ with decreasing $Ra_{c}(k)$ in a parametrically close
system of marginal-stability equations for which Eq. (\ref{eq:kmod})
holds. It could thus also apply to different problems,
where such formal conditions for the respective
stability parameter are met due to
other physical effects.

For a finite
$\theta>0$, the role of 3D disturbances is no longer
formally predictable. However, the above physical interpretation
of the effect of dissipation on the instability feedback still suggests
the 3D nature of the most unstable disturbances with relatively small
$k_{y}$. This effect is also due to the across-slot
gravity alone. It therefore has to vanish as the
slot orientation approaches vertical.

Examination of 3D oscillatory disturbances in the inviscid fluid [Figs.
\ref{f:mscoi}(b)---\ref{f:mscoi}(e)] shows that the disturbances with $k_{z}$
up to $3$ are more stable than the respective 2D perturbations ($k_{z}=0$).
Meaning that small-amplitude 3D disturbances of any physical nature are
not expected to arise in the range of parameters considered in Figs.
\ref{f:mscoi}(b)---\ref{f:mscoi}(e), this result is consistent
with the interpretation just mentioned. 

As also anticipated from the above interpretation, 3D oscillatory
disturbances are most unstable in viscous fluid for an interval of
$k_{y}$ adjoining $k_{y}=0$ [Figs. \ref{f:mscovs}(b)---\ref{f:mscovs}(e)
and \ref{f:mscovn}(b)---\ref{f:mscovn}(e)]. In addition, the data in Figs.
\ref{f:mscovs}(b)---\ref{f:mscovs}(e) and \ref{f:mscovn}(b)---\ref{f:mscovn}(e)
indicate that 3D disturbances are the first to arise only when the across-slot
gravity is present. [2D disturbances ($k_{z}=0$) are most unstable
for all $k_{y}$ presented in Figs. \ref{f:mscovs}(e) and \ref{f:mscovn}(e).]
Consistent with the above interpretation of the effect of the across-slot
gravity component, Figs. \ref{f:mscovs}(e) and \ref{f:mscovn}(e) thus
also suggest that the along-slot component does not introduce
an additional 3D oscillatory instability.\newline
\subsubsection{\label{s:rovv}Effect of the along-slot gravity.}
\paragraph{\label{s:rovvg}General.}
As the results in Figs. \ref{f:mscovs}(e) and \ref{f:mscovn}(e),
apart from the respective findings in \cite{r:trp}, suggest, 2D oscillatory
instability also arises due to the effect of the along-slot gravity component
alone. The mechanism of such instability can be identified by careful
examination of Fig. \ref{f:perv}. In this figure, temporal behavior
of an unstable perturbation mode possessing one of the reflectional
symmetries of the linearized problem is illustrated. 

The perturbation flow structures illustrated in Fig. \ref{f:perv}
suggest that their dynamics is underlain by two counter-propagating patterns.
Either such pattern is dominant near one of the slot sidewalls. The pattern propagating
in (against) the gravity direction, i.e. downwards (upwards), is dominant near the left
(right) sidewall in Fig. \ref{f:perv}. This could be viewed as manifestation of such
a perturbation being a superposition of two counter-propagating traveling waves. In
particular, let $q(\tau,x,y)$ stand for the reflectionally symmetric perturbation
of $\psi$, $t$, and $s$, nondimensionalized say consistently with Eqs.
(\ref{eq:ns1})---(\ref{eq:ds}). If $Ra_{c}(k)<Ra_{c}(nk)$ 
($n=2,3,...$), in the large-$\tau$ marginally
unstable state of such a wave number $k$, 
\begin{displaymath}
q(\tau,x,y)=
Re[a^{q}_{1}q'(x)e^{i(\omega_{c}\tau+ky)}+a^{q}_{2}q'(-x)e^{i(\omega_{c}\tau-ky)}]
\end{displaymath}
\begin{displaymath}
=Re\{a^{q}_{1}q'(x)e^{i(\omega_{c}\tau+ky)}
[1+\frac{a^{q}_{2}}{a^{q}_{1}}\frac{q'(-x)}{q'(x)}e^{-2iky}]\}
\end{displaymath}
\begin{equation}
=Re\{a^{q}_{2}q'(-x)e^{i(\omega_{c}\tau-ky)}
[1+\frac{a^{q}_{1}}{a^{q}_{2}}\frac{q'(x)}{q'(-x)}e^{2iky}]\}.
\label{eq:stw}
\end{equation}
Here coefficients $a^{q}_{1}$ and $a^{q}_{2}$ specify the symmetry of the
respective variable, and $q'(x)$ is the part of the \mbox{Fourier} mode of
the variable depending on $x$ alone. It is seen from Eq. (\ref{eq:stw})
that when $|q'(x)|\ll |q'(-x)|$ for $x>0$, the slot half near the left
(right) sidewall is dominated by the downwards-(upwards-)propagating
traveling wave. Dynamics of such a reflectionally symmetric mode
restricted to a slot half is thus largely due to the dominant
part of the respective propagating pattern.

Such dominant parts of the downwards- and upwards-propagating streamline
perturbation patterns are distinguishable as the along-slot sequences of counter-rotating
cells in the left and right halves, respectively, of Figs. \ref{f:perv}(d) and \ref{f:perv}(i).
Near the left (right) sidewall, the regions of the clockwise-rotating cells from such along-slot
cell sequence are seen from Fig. \ref{f:perv} to propagate downwards (upwards) concurrently
with mostly negative (positive) perturbation values of both components. The regions of such
counterclockwise-rotating cells are also seen to propagate downwards (upwards) concurrently
with mostly positive (negative) component perturbation values. Why such component
perturbation distributions take place is analyzed below.
\paragraph{\label{s:rovvd}Component perturbation distributions.}
Let us consider the dominant part of either traveling streamline perturbation pattern
[Figs. \ref{f:perv}(d) and \ref{f:perv}(i)] in the respective frame of reference moving with it.
In the reference frame moving with the downwards-(upwards-)propagating pattern, the laboratory-frame
velocities (Fig. \ref{f:perv}) oriented downwards (upwards) decrease. Such velocities
oriented upwards (downwards) thus increase. For an adequate speed of propagation, the component
perturbations are therefore transported vertically only against the direction of propagation
of the respective pattern. 

The distributions of temperature and solute perturbations 
in Figs. \ref{f:perv}(d) and \ref{f:perv}(i), in particular, are in
part the result of the vertical transport just highlighted. They are also
due to the advection associated with across-slot laboratory-frame velocities.
The orientation of such an advection is the result of addition of the along-slot
component of motion against the respective direction of propagation
to these laboratory-frame velocities. 

Allowing for the along-slot periodicity, the regions near the upper and
lower domain boundaries in Fig. \ref{f:perv}(d) are in the wake of the downwards-propagating
clockwise-rotating cells. Such a cell is seen in the left central part of the flow domain in
this figure. These regions are also in the wake of the upwards-propagating counterclockwise-rotating
cells represented by such cell in the right central part of the flow domain [Fig. \ref{f:perv}(d)].
The component perturbations are thus transported into the upper- and lower-boundary regions
by the velocities whose (laboratory-frame) orientation in such convective cells
is against the respective direction of propagation. 

The component perturbations in the upper- and lower-boundary regions in Fig.
\ref{f:perv}(d) are also affected via the advection caused by (laboratory-frame)
across-slot velocities. On addition of the along-slot component of motion against
the respective direction of propagation, such across-slot velocities in the
left (right) half of the slot are relevant only where they are oriented
in Fig. \ref{f:perv}(d) from (towards) the respective sidewall.

In all transport processes just outlined,
the temperature and solute perturbations are
transferred into the upper- and lower-boundary regions
[Fig. \ref{f:perv}(d)] only from the near-left-wall
and middle areas of the slot. In such areas, the
background values of the components are in the
upper half of their interval.

The central region in Fig. \ref{f:perv}(d), occupied by the full
counter-rotating cells, is also in the wake of two counter-propagating convective
cells. One of these is the upwards-propagating clockwise-rotating cell partly seen
in the right upper corner of the domain in Fig. \ref{f:perv}(d). The other is the
downwards-propagating counterclockwise-rotating cell partly seen in the left lower corner
of the domain. The component perturbations are thus transported into the central region of
the domain in Fig. \ref{f:perv}(d) by the velocities whose (laboratory-frame) orientation
in such convective cells is against the respective direction of propagation. Advection
due to the (laboratory-frame) across-slot velocities just below (above) the domain
central region in the left (right) slot half in Fig. \ref{f:perv}(d)
also affects the component perturbations in this region.

In all transport processes
just outlined, the temperature and solute perturbations are 
transferred into the central region of the domain in Fig. \ref{f:perv}(d) only from
the near-right-wall and middle areas of the slot. In such areas, the background
values of the components are in the lower half of their interval.

The central region of the domain in Fig. \ref{f:perv}(d) is thus
characterized by the relatively smaller values of the component perturbations while
the larger values are in the upper and lower regions of the domain. The dominant parts of
counter-propagating streamline perturbation patterns distinguished in Fig. \ref{f:perv}(i)
are about half of the wavelength ahead of their locations in Fig. \ref{f:perv}(d) in the
respective directions. The component perturbation distributions in Fig. \ref{f:perv}(i)
are consistent with the above understanding [described for Fig. \ref{f:perv}(d)]
of the transport processes given rise to by the cells in this figure. 

The component perturbations in Fig. \ref{f:perv}(i) can also be viewed as
resulting from their dominant parts being about half of the wavelength ahead of
their locations in Fig. \ref{f:perv}(d) in the respective directions. These dominant
parts are however indistinguishable in Figs. \ref{f:perv}(d) and \ref{f:perv}(i),
where they merge into the respective slot-wide patterns of the component
perturbations. They are only approximately discernible
in Figs. \ref{f:perv}(a) and \ref{f:perv}(e).

As co-rotating cells in the dominant parts of two counter-propagating streamline
perturbation patterns [Figs. \ref{f:perv}(d) and \ref{f:perv}(i)] have close
vertical locations, they merge into a single slot-wide cell. This is seen in Figs.
\ref{f:perv}(a)---\ref{f:perv}(c),\ref{f:perv}(e),\ref{f:perv}(f)---\ref{f:perv}(h),
and \ref{f:perv}(j). The temperature and solute perturbations in these figures result
from superposition of their respective counter-propagating perturbation patterns
[whose dominant parts are roughly discernible in Figs. \ref{f:perv}(a)
and \ref{f:perv}(e)]. The latter accompany the counter-propagating
streamline patterns [whose dominant parts are discerned in
Figs. \ref{f:perv}(d) and \ref{f:perv}(i)].

The component perturbation distributions in Figs.
\ref{f:perv}(a)---\ref{f:perv}(c),\ref{f:perv}(e),\ref{f:perv}(f)---\ref{f:perv}(h), and \ref{f:perv}(j)
are thus also consistent with the transport processes given rise to by the respective patterns of convective
cells in these figures. Such processes have to be interpreted as in the above analysis of Figs.
\ref{f:perv}(d) and \ref{f:perv}(i).
\paragraph{\label{s:rovvf}Feedback.}
In the pattern dominant part propagating downwards (upwards) near the left (right) sidewall,
the clockwise-rotating cells thus steadily coincide with the regions of mostly negative
(positive) perturbation values of the components. Such counterclockwise-rotating cells
then steadily coincide with the regions of mostly positive (negative) component
perturbation values [Figs. \ref{f:perv}(d) and \ref{f:perv}(i)]. In the vicinity
of a sidewall, the perturbation isotherms are however prone to be parallel to
the boundary, where the temperature perturbation vanishes. In contrast to
this, the solute perturbation isolines tend to be orthogonal to the
sidewalls, because of the flux-free boundary conditions.

Such disparate isoline behavior gives rise to differential gradient diffusion. As a
consequence, a near-slot-middle streamline point in the downwards-(upwards-)propagating
clockwise-rotating cells [Figs. \ref{f:perv}(d) and \ref{f:perv}(i)] is typically heavier
(lighter) than the respective streamline point near the left (right) sidewall at the same vertical
location. Likewise, a near-slot-middle streamline point in the downwards-(upwards-)propagating
counterclockwise-rotating cells is typically lighter (heavier) than the respective streamline
point near the left (right) sidewall at the same vertical location. [In particular,
see Figs. \ref{f:pervd}(d) and \ref{f:pervd}(i).] 

The flow perturbation structures illustrated in Fig. \ref{f:perv} are expected to
result from superposition of the counter-propagating patterns whose dominant parts have
just been described. They are all thus characterized by a horizontal density difference
between two streamline points that is consistent with the sense of rotation of the
respective convective cell. This can be inferred from Fig. \ref{f:perv} and
is explicitly illustrated in Fig. \ref{f:pervd}.

In a quasisteady sense, therefore, the energy that drives
small-amplitude oscillatory convection in Fig. \ref{f:perv} has essentially the
same nature as that driving small- and finite-amplitude steady convection in
\cite{r:tlh,r:trp}. In contrast to small-amplitude steady convection, such
oscillatory convection arises for $\mu=1$ because the oscillatory perturbation
gives rise to a more favorable ratio between the maximal absolute values of
the relative perturbations of the components. This is discussed in Sec.
\ref{s:rslv} below, in the context of steady convection.

The streamline disturbance resulting from superposition of certain
counter-propagating convective flow patterns thus generates such distribution
of the component perturbations as favors its convective motion. For amplitude
growth of such a disturbance to be maintained, however, the counter-propagation
of the convective patterns also has to be sustained. This underlies
the oscillatory nature of the instability.

It was assumed above that, in the marginally unstable state,
the speed of streamline-pattern propagation adequately matches
the laboratory-frame convective velocities oriented in the
respective direction of propagation. The corresponding
streamline points thus have their component
perturbation values practically steady.

In the counter-propagating sequences of convective cells
[Figs. \ref{f:perv}(d) and \ref{f:perv}(i)], practically steady
values of the component perturbations arise in the near-wall regions
of the counterclockwise-rotating cells. The values of component perturbations in
these cell sequences also have to be practically steady in the middle slot regions
of the clockwise-rotating cells. For all these regions, the laboratory-frame velocities
are oriented in the direction of propagation of the respective streamline pattern.
In the middle slot regions, the effects of temperature and solute perturbations
on the density perturbation largely offset each other. Due to differential
gradient diffusion, however, the density perturbation in the near-wall
regions is specified primarily by the respective values
of solute perturbation alone.

The solute (density) perturbation values in the near-wall regions of
counterclockwise-rotating cells in Figs. \ref{f:perv}(d) and \ref{f:perv}(i)
[Figs. \ref{f:pervd}(d) and \ref{f:pervd}(i)] are expected to be steadily
positive and negative in the left and right parts of the slot, respectively.
The time instances when such counter-propagating counterclockwise-rotating
cells merge into a single counterclockwise-rotating cell are illustrated in Figs.
\ref{f:perv}(a)---\ref{f:perv}(c),\ref{f:perv}(e),\ref{f:perv}(f)---\ref{f:perv}(h),
and \ref{f:perv}(j). The values of solute (density) perturbation in such a
slot-wide cell near the left sidewall are also always larger than
those near the right sidewall (Fig. \ref{f:pervd}).

It is the steady horizontal density difference arising between two halves of the
slot that sustains the counter-propagation underlying dynamics of the reflectionally
symmetric disturbance. The counter-propagating convective streamline patterns
thus give rise to such feedback distribution of the component perturbations
as, due to differential gradient diffusion, enhances their convective
motion and sustains their counter-propagation.
\paragraph{\label{s:rovvm}Marginal-stability boundaries.}
Differential gradient diffusion is most effective when a convective cell
particle has the maximal time available at a particular wavelength to unequally change
its component values while its across-slot location is changed. This time is the larger
the longer the wavelength is. In order that the oscillatory instability mechanism just
described be effective, such diffusion time has to be adequately matched by the time
during which the wavelength is passed by either traveling perturbation pattern.
The latter time specifies the marginal instability frequency. Such frequency
in Figs. \ref{f:mscovs}(e) and \ref{f:mscovn}(e) thus decreases as the
wavelength increases. (In this Sec. \ref{s:rovvm},
$k_{z}=0$ and thus also $k=k_{y}$ are implied.)

Dissipation of the along-slot component of perturbation motion is
part of the feedback for the steady instability caused by different boundary
conditions in a vertical-slot geometry (see Sec. \ref{s:rslv} below).
The efficiency of such a feedback thus has to decrease as the wavelength increases,
as discussed for the horizontal-slot ($\theta=0$) oscillatory instability
feedback (Sec. \ref{s:rovh}).

In view of the vertical-slot oscillatory instability feedback being
of a quasisteady nature, its efficiency is also progressively reduced
by the dissipation of along-slot perturbation motion when the wavelength
increases. As in the horizontal-slot oscillatory instability, the feedback
efficiency would also decrease to zero if the time available for across-slot
diffusion as $\lambda\rightarrow\infty$ ($k\rightarrow 0$) were fully utilized.
In the long-wavelength limit, therefore, $\omega_{c}(k)$ in Figs. \ref{f:mscovs}(e)
and \ref{f:mscovn}(e) remains a finite nonzero value. However, differential
gradient diffusion, being a basic ingredient of the vertical-slot oscillatory
instability mechanism, would then have to progressively decrease its effectiveness as
$k$ approaches $0$. This explains why $Ra_{c}(k)$ increase as $k$ decreases near $k=0$
in Figs. \ref{f:mscovs}(e) and \ref{f:mscovn}(e) and the (2D) marginal-stability curves
in these figures [as well as those in Figs. \ref{f:mscovs}(b)---\ref{f:mscovs}(d)
and \ref{f:mscovn}(b)---\ref{f:mscovn}(d)] remain qualitatively the
same as in Figs. \ref{f:mscovs}(a) and \ref{f:mscovn}(a).

As $\theta$ increases above $\pi/2$, the emergence of the across-slot
gravity component makes the associated mechanism of linear steady instability
\cite{r:twh,r:tbc} potentially relevant. For a particular wavelength, the diffusion
time thus has to optimally match a combination of the steady manifestation of the effect
of the across-slot gravity component and the oscillatory one of the along-slot component.
The marginally unstable frequency, $\omega_{c}(k)$, is thus generally expected to
decrease as $\theta$ exceeds $\pi/2$. As the wavelength increases, however, such frequency
specified by the effect of the along-slot gravity component alone progressively exceeds
that corresponding to the maximal time available at the wavelength for the respective
manifestation of differential gradient diffusion. This allows to avoid
the overly large reduction of the feedback efficiency.

So long as the effect of the along-slot gravity component is dominant, 
the dissipation of along-slot perturbation motion remains of primary importance.
Intermediate and long wavelengths could not thus significantly decrease their marginally
unstable frequency. For this reason, $\omega_{c}(k)$ decreases noticeably only in the
region of relatively large $k$ in Figs. \ref{f:mscovs}(f) and \ref{f:mscovn}(f), whereas
it remains practically unchanged in the vicinity of $k=0$. The intermediate and long
wavelengths thus fail to substantially absorb the steady manifestation of the effect of
the across-slot gravity component. It is only for the relatively large wave numbers,
therefore, that $Ra_{c}(k)$ decrease with increasing $\theta$. In contrast to this,
the decreasing effect of the along-slot gravity component leads even to slight
increases of $Ra_{c}(k)$ in the interval of relatively small wave numbers
in Figs. \ref{f:mscovs}(f) and \ref{f:mscovn}(f).
 
With $\theta$ increasing above $\pi/2$, the decrease of the along-slot gravity component
still enhances room for the growing across-slot component. The lower bound of the region
of $k$ with decreasing $Ra_{c}(k)$ is thus found to decrease [Figs. \ref{f:mscovs}(f)
and \ref{f:mscovn}(f)]. Around $\theta\approx1.5\pi/2$, however, $Ra_{c}(k)$ are still
increasing (with increasing $\theta$) when $k\leq\sim2.5$ for both stress-free and no-slip
boundary conditions. In particular, the development of a new minimum in the shape of $Ra_{c}(k)$
[Figs. \ref{f:mscovs}(f) and \ref{f:mscovn}(f)] is a manifestation of the opposing trends
exhibited by $Ra_{c}(k)$ within the different parts of the interval of $k$. 

For $\theta\approx1.55\pi/2$, $\omega_{c}(k)$ in the regions of the smallest and
largest values of $k$ in Figs. \ref{f:mscovs}(f) and \ref{f:mscovn}(f) could not
be continued into each other. Slope $|\partial\omega_{c}(k)/\partial k|$ also
seemed to exhibit infinite increases around an interval of middle values of
$k$. No marginally unstable curve with $\omega_{c}(k)>0$ was then found
within this interval of middle $k$. These findings suggest that at
$\theta\approx$ ($1.5$---$1.55$)$\pi/2$ the oscillatory marginal-stability
boundaries in Figs. \ref{f:mscovs}(f) and \ref{f:mscovn}(f)
begin to transform into the respective curves
characterized by a zero real eigenvalue.
\section{\label{s:rs}Steady convection}
\subsection{\label{s:rsl}Linear steady instability}
\subsubsection{\label{s:rslv}Effect of the along-slot gravity.}
The vertical slot with inviscid fluid was found in Sec.
\ref{s:roiv} above to be unstable to steady infinitesimal
disturbances for a continuum of wave numbers at any $Ra\neq 0$. In viscous fluid,
however, dissipation of the along-slot component of perturbation-cell motion is
part of the feedback for steady instability, if any, in the vertical-slot geometry.
In particular, the component gradient disparity and resulting differential gradient
diffusion are caused by across-slot perturbation motion. Such across-slot motion
is however maintained only due to along-slot motion in the convective cell.
Driven by the horizontal density differences between respective
streamline points, such along-slot cell motion arises from
differential gradient diffusion \cite{r:tlh,r:trp},
the process responsible for the feedback. 

For viscous fluid, the linear steady instability of all wavelengths is thus generally
less likely in a vertical-slot geometry than in the horizontal slot where the component
stratification favors steady instability \cite{r:twh,r:tbc}. In addition, if such a vertical-slot
instability does arise (as in \cite{r:tlh}, for example, where the background horizontal
gradients are not exactly compensating), the long wavelengths are not expected to be most
unstable. This is associated with the feedback efficiency tending to zero as $\lambda\rightarrow\infty$.
As seen from Figs. \ref{f:mscs} and \ref{f:mscn}, however, the long wavelengths are most unstable
for all illustrated $\theta$. Their instability is also most persistent as the interval
of unstable wave numbers shrinks and vanishes when $\theta\rightarrow\pi/2$. Even when
$\theta$ is very close to $\pi/2$, therefore, the instability could come only
from the effect of the across-slot gravity component. This effect is known
to result in such a long-wavelength instability \cite{r:twh,r:tbc,r:nllm}.

The viscous manifestation of linear steady instability in a
vertical-slot geometry seems to be limited to a more favorable background
ratio between the opposing component gradients. Indeed, enhancement of the relative
background solute scale with respect to that in a LHSSS ($Ra_{S}>0$, $\mu=0$) \cite{r:trp},
taking place as $\mu$ increases from $0$, is expected to be unfavorable for development
of the LHSSS linear steady instability. In particular, the relative increase of the solute
perturbation amplitude, associated with such solute scale enhancement, would result in a
decrease of the horizontal density differences between respective streamline points of
the perturbation pattern illustrated in Fig. 2(a) of \cite{r:tlh}. This was
confirmed in trial computations for $Ra_{S}=30000$ and $\lambda=2$. 

Essentially the same findings are reported for $Ra_{S}=0$
as well (Fig. \ref{f:permu}), when $\mu$ increases from $\sim0.92$ to
$\sim0.98$. The streamline perturbation pattern in Fig. \ref{f:permu}(a),
which is similar to that in Fig. 2(a) of \cite{r:tlh}, could not be maintained as the
relative solute perturbation amplitude grows when $\mu$ is increased towards $1$. Such
pattern is thus found to undergo changes [Figs. \ref{f:permu}(b) and \ref{f:permu}(c)].
The changed streamline pattern in Fig. \ref{f:permu}(c), however, only
enhances the unfavorable role of dissipation, which results
in vanishing of the instability.

The reported analysis of linear steady instability for $Ra_{S}=0$ was possible in
view of the existence of a narrow interval of $\mu$ ($\sim0.86\leq\mu\leq\sim0.98$)
for which $\lambda=2$ is unstable to steady infinitesimal disturbances. This interval has
been uncovered in the present study. When $\mu$ is decreased below $\sim0.86$, the linear
steady instability disappears. Vertical shear motion, intensifying with decreasing $\mu$,
becomes an important factor in the nature of the background flow. This prevents the small-amplitude
manifestation of differential gradient diffusion. As $\mu$ exceeds $\sim0.98$, on the
other hand, the ratio between the background component scales becomes insufficiently
favorable for the manifestation of small-amplitude steady convection. This ratio
matters so much because of the effect of dissipation
on the instability feedback.

The quasisteady nature of oscillatory instability in the vertical-slot geometry
suggests that the relative values of the component perturbations are critically
important for the manifestation of such instability as well. In particular, the streamline
patterns in Figs. \ref{f:perv}(c),\ref{f:perv}(g), and \ref{f:perv}(h) are reminiscent of
those in Fig. \ref{f:permu}. The maximal absolute values of solute perturbation in Figs.
\ref{f:perv}(c),\ref{f:perv}(g), and \ref{f:perv}(h) are however smaller, compared to the
respective temperature perturbation values, than those in Fig. \ref{f:permu}. Counter-propagation
of the streamline disturbance constituents prevents such disturbance from coherently maintaining
the maximal absolute values of the solute perturbation exceeding those of the temperature
perturbation. The dominant part of either traveling streamline perturbation pattern also
interacts only with one of the sidewall boundary conditions. In contrast to this,
the steady disturbance interacts with both such conditions. 
\subsubsection{\label{s:rslh}Effect of the across-slot gravity.}
When $Ra>0$ and $\theta=\pi$, different boundary conditions result in steady
linear instability. The mechanism of this instability \cite{r:twh,r:tbc} has
to be entirely responsible for small-amplitude steady convection for
$\theta\in(\pi/2,\pi)$ as well (Sec. \ref{s:rslv}).

At any $\theta\in(\pi/2,\pi)$ illustrated in Figs. \ref{f:mscs} and \ref{f:mscn},
the instability is restricted to the wave numbers that are smaller than a certain
critical value. From the qualitative point of view, this is consistent with enhancement
of the overall dissipation by the wave number increase. However, such largest unstable
wave number decreases to zero as $\theta$ approaches $\pi/2$. This does not seem to be
a necessary consequence only of the vanishing of the across-slot gravity component.
In addition, the increases of $Ra_{c}(k)$ in the vicinity of the largest unstable
$k$ for a given $\theta$ in Figs. \ref{f:mscs} and \ref{f:mscn} are noticeably
more abrupt than one could expect from the behavior of $Ra_{c}(k)$ at the same
$k$ and larger $\theta$. The behavior of $Ra_{c}(k)$ is thus affected
by an additional stabilizing factor.

The curves in Figs. \ref{f:mscs} and \ref{f:mscn}
are affected by the along-slot gravity component. In the presence of this
gravity component, the steady perturbation maintained by the across-slot component would
have to change its pattern. However, the along-slot gravity component alone fails to give
rise to linear steady instability. Such perturbation change could thus be only stabilizing,
for the resulting pattern would no longer be optimal with respect to the effect of the
across-slot gravity component. In particular, the perturbation change could result in
the across-slot cell path acquiring a slope with respect to the
across-slot coordinate. This would enhance the dissipation of across-slot
motion. Such stabilizing effect is also most pronounced for small
wavelengths, and its relative significance decreases
as the wavelength increases.

Another feature of Figs. \ref{f:mscs} and \ref{f:mscn} is that the interval of
unstable wave numbers is longer for no-slip boundary conditions (Fig. \ref{f:mscn})
when $1.4\pi/2\leq\theta<\pi$. For $\pi/2<\theta\leq1.3\pi/2$, it is the stress-free
conditions (Fig. \ref{f:mscs}) that give rise to a longer interval of unstable wave
numbers. In the latter case, all unstable wavelengths are large enough for the
dissipation of along-slot perturbation motion to dominate the overall dissipation.
The stress-free boundary conditions thus allow to maintain a longer interval of unstable
wavelengths. As the slot orientation is far enough from vertical, however, the
shortest unstable wavelengths are such that the dissipation of across-slot
motion becomes relevant. The respective critical component
gradients are then larger for no-slip boundary conditions. This factor
proves to be more important for maintaining the instability of the
shortest wavelengths than the dissipation decrease
due to the stress-free boundary conditions.

To obtain the dependence of $Ra_{c}(0)$ on $\theta$, 
\begin{displaymath}
Ra_{c}(k)=Ra_{0}+Ra_{2}k^{2}+...,
\end{displaymath}
\begin{displaymath}
\tilde{u}(k)=k^{2}(\tilde{u}_{0}+\tilde{u}_{2}k^{2}+...),
\end{displaymath}
\begin{displaymath}
\tilde{t}(k)=\tilde{t}_{0}+\tilde{t}_{2}k^{2}+...,
\end{displaymath}
\begin{equation}
\tilde{s}(k)=\tilde{s}_{0}+\tilde{s}_{2}k^{2}+...  \label{eq:lwes}
\end{equation}
were introduced into Eqs. (\ref{eq:lssu})---(\ref{eq:lsss}). $\tilde{t}_{0}=0$ and
$\tilde{s}_{0}=const$ were thus obtained from the $k^{0}$ order of Eqs. (\ref{eq:lsst})
and (\ref{eq:lsss}) and the boundary conditions for $\tilde{t}_{0}$ and $\tilde{s}_{0}$.
Such $\tilde{t}_{0}$ and $\tilde{s}_{0}$ were then introduced into the $k^{1}$ order of
Eq. (\ref{eq:lssu}). Along with either (stress-free) boundary conditions (\ref{bc:vs})
or (no-slip) boundary conditions (\ref{bc:vn}), this resulted in the respective
expressions for $\tilde{u}_{0}$. With the use of these expressions, the
$k^{2}$ order of Eq. (\ref{eq:lsss}) and the respective boundary
conditions for $\tilde{s}_{2}$ lead to 
\begin{equation}
Ra_{c}(0)=Ra_{0}=-Ra_{c}^{\pi}/\cos\theta=Ra_{c}^{\pi}/\sin(\chi\pi/2). \label{eq:racs}
\end{equation}
Here $Ra_{c}^{\pi}$ is the value $Ra_{c}(0)$ would take on in the horizontal slot
($\theta=\pi$): $Ra_{c}^{\pi}=120$ for stress-free boundary conditions and
$Ra_{c}^{\pi}=720$ for the no-slip conditions \cite{r:nllm}. Such values
were first encountered in \mbox{Rayleigh---Benard} convection
with the flux unstable gradient \cite{r:sh}.

The derivation of Eq. (\ref{eq:racs}) shows that the terms
associated with the along-slot gravity component in the right-hand side of
Eq. (\ref{eq:lssu}) have no impact on Eq. (\ref{eq:racs}). This is consistent
with the above interpretations that the effect of this gravity component is
not among the causes of linear steady instability for $\theta\in(\pi/2,\pi)$
and that any role of this effect vanishes as $k\rightarrow0$.
Eq. (\ref{eq:racs}) was also found to accurately coincide
with the numerical data underlying Figs.
\ref{f:mscs} and \ref{f:mscn}.
\subsection{\label{s:rsn}Finite-amplitude steady convection}
\subsubsection{\label{s:rsng}General.}
The specification of solute concentration at the middle points of the across-slot
boundaries was found to numerically permit two pairs of periodic steady convective
solutions arising from the conduction state. Within the selected along-slot period,
one of such solution pairs is characterized at the onset of linear instability by
two full counter-rotating cells. Except for the horizontal-slot geometry, the
corresponding finite-amplitude solutions are generally reflectionally asymmetric.
This is the result of an asymmetry between their counter-rotating
cells (see Sec. \ref{s:rsns} below).

In the other solution pair, a clockwise-(counterclockwise-)rotating
cell in the center of the period interval is combined with parts
of counterclockwise-(clockwise-)rotating cells near the across-slot
boundaries. This pair possesses the reflection symmetry identified by
$LS[\psi(-x,-y),t(-x,-y),s(-x,-y)]^{T}=[\psi(x,y),-t(x,y),-s(x,y)]^{T}$
for all $\theta$. The eigenvectors in Fig. \ref{f:permu}, in particular,
correspond to the solutions that are $LS$-symmetric
for the illustrated along-slot interval.

The only essential difference between the branch pairs is
however the along-slot phase shift between their solutions.
In particular, when linear steady instability is present for
a given $\theta$ (say for $\theta=\pi$ or for $\theta=0$ as
$\mu<1$), these solution pairs bifurcate from the conduction
state at the \mbox{Rayleigh} numbers the difference between which
is practically negligible. When the bifurcation is subcritical, the
limit points are also formed at practically the same $Ra$. These
pairs of branches can thus be viewed as standing for a single
physically relevant pair of solution branches arising from
the respective conduction state.

The two pairs of convective solution branches were however found to undergo
dissimilar secondary steady bifurcation phenomena, as $Ra$ or $\theta$ were varied.
Such phenomena were detected by changes in the solution Jacobian sign, monitored
during the continuation procedure. This dissimilarity could arise, in particular,
from the different symmetries of the respective convective solutions, only one pair
of which is $LS$-symmetric. In terms of the flow structure, however, the encountered
secondary bifurcation phenomena were often found to give rise only to along-slot
phase shifts of the respective convective patterns. The presence or absence of
secondary steady bifurcations, and of the instabilities associated with
them, in real fluid could not thus be established with certainty
in the framework of the present numerical approach.

In addition, the stability of finite-amplitude steady
convective branches to oscillatory disturbances was found to depend
on which of the two numerically present pairs of solutions is examined.
This proved to hold even for the horizontal-slot geometry, where
the flow pair with full counter-rotating cells also
has a reflection symmetry.

The prescribed across-slot-boundary values of $s$ can
prevent traveling-wave disturbances from being detected in the present
formulation. However, finite-amplitude traveling-wave branches could intersect
steady convective branches and thus affect the stability of the latter. Such findings
have been previously reported for binary fluid convection \cite{r:blks} and for
conventional double-diffusive convection \cite{r:dkt}. The presence of traveling-wave
branches also affects the stability of reflectionally symmetric oscillatory branches
\cite{r:ckgs}. This could have an additional effect on the stability of
finite-amplitude steady solutions. The stability of steady convective
branches to secondary oscillatory disturbances may therefore anyway
change if the existence of traveling waves
is properly allowed for.

The solution structures reported below thus refer to the two pairs of steady
convective branches arising in such numerical computations as one, as has been done
for the periodic solution computations in \cite{r:tlh,r:trp,r:tpla} as well. The
focus of this study is also on the physical mechanisms of conduction-state
instability. For the reasons just mentioned, therefore, the issues of
secondary (steady and oscillatory) instability of the computed
finite-amplitude steady flows were left outside
the scope of this work.
\subsubsection{\label{s:rsns}$\theta\in[\pi/2,\pi]$.}
For a large enough perturbation amplitude at $\theta=\pi/2$, an appropriate
streamline pattern is maintained by the combined action of differential gradient
diffusion and along-slot gravity \cite{r:trp}. This is possible despite the
effects of dissipation, due to the energy coming from the nonlinear
terms in Eqs. (\ref{eq:ns1})---(\ref{eq:ds}).

As $\theta$ changes from $\pi$ to $\pi/2$, the mechanism of differential gradient
diffusion discussed in \cite{r:twh,r:tbc} thus gradually gives way to that considered
in \cite{r:trp}. Via supercritical bifurcation, the former mechanism gives rise to a
convective steady pattern of symmetric counter-rotating cells. The steady convection
pattern resulting from the finite-amplitude mechanism in \cite{r:trp} has however
only clockwise-rotating cells. The along-slot gravity component has an asymmetric
effect on the senses of rotation of the counter-rotating cells.

In particular, let a perturbation pattern of symmetric counter-rotating cells be
generated in a vertical slot with two background horizontal gradients and the higher
component values at the left sidewall. The clockwise-rotating cells would then create locally
unstable solute perturbation stratifications. Such stratifications would however be stable where
the cells rotate counterclockwise. Due to differential diffusion (whether of gradient or
classical nature), the clockwise-rotating cells have to intensify with respect to
the cells rotating counterclockwise \cite{r:tlh,r:ths,r:trp,r:xqt}.
The clockwise-rotating cells are thus dominant in
the finite-amplitude steady pattern.

At the onset of linear steady instability, the counter-rotating cells are bound
to be symmetric. The asymmetry between them could thus arise, as $\theta$ changes
from $\pi$ to $\pi/2$, only with a finite convection amplitude. The latter is expected
to arise when the \mbox{Rayleigh} number increases above the linear instability onset. The
effect of the along-slot gravity component is itself also enhanced as the \mbox{Rayleigh}
number is increased. For $Ra=31000$ and $\lambda=2$, in particular, the flow patterns are
thus entirely dominated by the clockwise-rotating cells when $\theta$ is still greater
than $1.9\pi/2$ [Figs. \ref{f:nonsm}(a)---\ref{f:nonsm}(d)]. The remnants of the
counterclockwise-rotating cells also vanish for $\theta\in(1.7\pi/2,1.75\pi/2)$
[Figs. \ref{f:nonsm}(e)---\ref{f:nonsm}(h)].

For $\theta\in[\pi/2,\pi]$, differential gradient diffusion underlies the
convective mechanisms resulting from both the along-slot and the across-slot
gravity components. The mechanism at $\theta=\pi/2$, however, leads only to
nonlinear steady instability. As $\theta$ changes from $\pi$ to $\pi/2$, primary
control over finite-amplitude steady convection is thus continuously transformed
from the convective mechanism arising from the across-slot gravity component to that
associated with the along-slot component. Such linear instability can however be due to the
decreasing across-slot gravity component alone. At some $\theta[$ $\approx$ ($1.6$---$1.7$)$\pi/2$
for $\lambda=2$], this renders the steady bifurcation subcritical, as seen from Figs.
\ref{f:bdra}(a) and \ref{f:bdra}(b). As the across-slot gravity component further
decreases with $\theta$ being decreased [to $\theta\in(1.3\pi/2,1.4\pi/2)$
for $\lambda=2$], the linear steady instability vanishes
[Figs. \ref{f:bdra}(c) and \ref{f:bdra}(d)]. 
\subsubsection{\label{s:rsno}$\theta\in[0,\pi/2]$.}
For $\theta\in[0,\pi/2]$, linear steady instability can arise
for $\mu=1$ neither from the effect of the along-slot gravity component
nor from that of the across-slot component. The latter effect favors amplitude
growth only of the small-amplitude convective cells whose sense of rotation changes
periodically in time (Secs. \ref{s:roih} and \ref{s:rovh}).
In addition, differential gradient diffusion is not among the causes of
finite-amplitude steady convection at $\theta=0$. Rather, it plays only
a stabilizing role in the mechanism of such convection \cite{r:tpla}.

The nature of finite-amplitude steady convective flows at $\theta=0$ is thus basically
different from that at $\theta=\pi/2$. As the convection amplitude becomes large enough,
with the \mbox{Rayleigh} number being increased, such dissimilarity in the nature of
convection prevents continuous transformation of one of the convective steady flows
into the other. In particular, a region of hysteresis in $\theta$
is formed for $Ra>\sim20000$ (Fig. \ref{f:bdthet}).

As $\theta$ increases from $0$, finite-amplitude steady convection at
$Ra=31000$ remains under primary control of the effect of the across-slot
gravity component. Playing a stabilizing role in this effect \cite{r:tpla},
differential gradient diffusion reduces the relative significance of its
destabilizing role in the emerging effect of the along-slot gravity component.
The counterclockwise-rotating cells thus retain their horizontal-slot pattern
along the whole branch $A2_{h}$ [Figs. \ref{f:nons}(a) and \ref{f:nons}(b)].
For $Ra=31000$ and $\lambda=2$, this branch extends to about
$\theta\approx0.51\pi/2$ [Fig. \ref{f:bdthet}(c)].

With $\theta$ decreasing from $\pi/2$, finite-amplitude steady
convection remains under primary control of the effect of the along-slot gravity
component. This effect is underlain by differential gradient diffusion \cite{r:tlh,r:trp}.
Such process reduces the effectiveness of the convective mechanism by means of which the
emerging across-slot gravity component acts \cite{r:tpla}. This gravity component thus fails
to have a significant effect on the convective pattern (i.e., to generate counterclockwise-rotating
cells). The whole branch $A2_{v}$, extending to $\theta\approx0.38\pi/2$ for $Ra=31000$
and $\lambda=2$, is therefore characterized only by clockwise-rotating cells [Fig.
\ref{f:nons}(d)]. Along branch $A2_{hv}$ [Fig. \ref{f:bdthet}(c)], such
convection pattern is transformed into that with counter-rotating
cells [Figs. \ref{f:nons}(b) and \ref{f:nons}(c)].
\section{\label{s:c}Summary and concluding remarks}
This work provides a comprehensive understanding of the manifestation of double-component
convection due to the effects of different component boundary conditions in an infinite
slot having diverse orientations to the gravity. To eliminate other physical effects, the
primary focus has been on the problem with compensating gradients of the components and
equal component diffusivities. Small-amplitude oscillatory convection as well as small-
and finite-amplitude steady convection have been considered. Although the study was
primarily focused on 2D convection, small-amplitude 3D effects were also
examined where the obtained physical insight
suggested their potential relevance.

The main aspect of the provided understanding is a consistent interpretation of the
physics underlying small-amplitude oscillatory convection. For all slot orientations,
the physical nature of such convection both in inviscid and in viscous fluid proved
to be underlain by differential gradient diffusion. However, the specific mechanisms
by means of which the manifestation of this effect takes place vary significantly
with the slot orientation. Their features also depend substantially on
whether the involved fluid is of inviscid or viscous nature.

In the horizontal slot at $\theta=0$ with inviscid fluid and an independently
prescribed flux stratification, $Ra^{s}$, the marginal-stability
boundaries behave consistently with the physical mechanism of oscillatory instability
first highlighted in \cite{r:wel}. In particular, the most unstable wave number is
zero and the instability arises within a finite wave number interval. The oscillation
frequency also decreases with decreasing wave number and becomes zero at $k=0$.
By means of the long-wavelength expansion, the critical fixed-value
stratification, $Ra_{c}$, and the group velocity,
$\partial\omega_{c}/\partial k$, for $k=0$ were
expressed in terms of $Ra^{s}$.

For the compensating background gradients in inviscid fluid, the most
unstable wave number remains equal to zero up until the slot orientation becomes nearly vertical
($\theta=\pi/2$). The analytical relation between $Ra_{c}$ and $\partial\omega_{c}/\partial k$
at $k=0$ was derived for such inclined slot at any $\theta\in[0,\pi/2)$ as well. For $\theta$
growing from $0$ to $\pi/2$, the oscillatory instability is however increasingly affected
by the along-slot gravity component. The effect of this gravity component is associated
with the mechanism of steady instability identified in \cite{r:tlh,r:trp} for viscous
fluid. The oscillatory marginal-stability curves for $\theta\in(0,\pi/2)$ were found
to behave consistently with the assumption that the corresponding mechanism in
inviscid fluid is effective at any $Ra\neq 0$. In particular, the oscillatory
marginal-stability boundary was found to transform into the steady
one with $Ra_{c}(k)=0$ as $\theta\rightarrow\pi/2$. 

In the horizontal slot ($\theta=0$) with viscous fluid,
the vicinity of zero wave number is no longer the region of most unstable $k$. 
$\omega_{c}(k)$ also tends to a nonzero value as $k\rightarrow0$. These findings
are still consistent with the interpretation of the oscillatory instability
for inviscid fluid. In particular, the dissipation of along-slot perturbation
motion was found to be part of the feedback in the horizontal-slot oscillatory
instability mechanism. Increase of the instability wavelength thus
reduces the efficiency of such feedback. As $k$ approaches $0$,
this stabilizing factor becomes more important than the
destabilizing decrease of the overall dissipation.

The suggested interpretation of the role played by the dissipation of along-slot
motion in the horizontal-slot oscillatory instability feedback
apparently applies to the classical double-diffusion as well. In particular,
zero wave number has to be most unstable in the diffusive regime of conventional
double-diffusive convection when both background gradients in inviscid fluid
are maintained by flux boundary conditions. Indeed, Fig. 4(b) in \cite{r:mvh}
seems to suggest that the most unstable wavelength tends to infinity and the
oscillation frequency tends to zero as $Pr\rightarrow0$. For a finite $Pr$,
however, both the most unstable wavelength and the corresponding
oscillation frequency in Fig. 4(b) of \cite{r:mvh} are finite.

For relatively large $\lambda_{y}$, most unstable disturbances of the
oscillatory instability in an inclined slot [$\theta\in(0,\pi/2)$], where
the across-slot gravity component is involved in driving the instability,
are of a 3D nature. This was also suggested to result from the effect of
dissipation of along-slot perturbation motion. In particular, the feedback
efficiency increases when $\lambda$ decreases due to orientation change of the
2D axis of rotation of a perturbation cell. All relevant findings proved to be
consistent with such interpretation of the 3D instability nature. For small
$\theta>0$, this interpretation was also found to be a manifestation of
more general conditions for three-dimensionality of instability.
Such conditions were formulated independently of the
physics underlying a specific problem.

Another important outcome of this study is identification of the
oscillatory instability mechanism arising from the effect of the along-slot
gravity component in viscous fluid. The reflectionally symmetric perturbation
mode of this instability consists of two counter-propagating patterns either of
which is dominant near one of the slot sidewalls. The counter-traveling convective
streamline patterns give rise to such feedback distribution of the component perturbations
as generates their amplitude growth and a nearly steady density difference across
the slot. This density difference sustains the counter-propagation of the convective
patterns. Such growth of the convection amplitude and formation of the across-slot
density difference are underlain by differential gradient diffusion. In a
quasisteady sense, the manifestation of such differential diffusion
is analogous to that in the respective steady
instability \cite{r:tlh,r:trp}. 

The identified mechanism of vertical-slot oscillatory instability
is apparently also relevant under the conditions other than those
considered in this work. In particular, Fig. 5 in \cite{r:m} suggests
that oscillatory instability arises in the vertical slot filled with a
porous medium even when different boundary conditions could be the only
cause of this instability. In addition, the instability mechanism identified
in the present study may also be effective when the role of differential
gradient diffusion in it is played by the classical double-diffusion. Indeed,
the directions in which counter-propagating flow patterns travel in Fig. 8
of \cite{r:m} seem to be consistent with the respective directions in Fig.
\ref{f:perv} of the present work. The eigenfunctions of traveling waves
illustrated in Fig. 7 of \cite{r:m} are also dominated by
their patterns near the respective slot boundaries.

The quasisteady nature
of oscillatory instability in the vertical slot
suggests that the dissipation of along-slot perturbation motion
is part of the instability feedback, as also highlighted for the
respective mechanism of steady instability. Since such a dissipation
is part of the oscillatory instability feedback at $\theta=0$ as well,
the same shape of the oscillatory marginal-stability boundaries for all
$\theta\in[0,\pi/2]$ is thus explained. In contrast to small-amplitude
steady vertical-slot convection, the manifestation of such quasisteady
oscillatory convection is permitted due to a relatively more
favorable ratio between the maximal absolute values of
the component perturbations.

As $\theta$ exceeds $\pi/2$, the mechanism of linear steady
instability arising from the effect of the across-slot gravity component
\cite{r:twh,r:tbc} becomes increasingly involved in the manifestation of
oscillatory instability due to the along-slot component. Behavior of the
oscillatory marginal-stability curves above $\theta\approx1.5\pi/2$ was
thus found to suggest their transformation into the respective
curves characterized by a zero real eigenvalue. 

The effect of the across-slot gravity component identified in \cite{r:twh,r:tbc} 
is the only cause of linear steady instability for $\theta\in(\pi/2,\pi]$. Driven
by differential gradient diffusion, such instability arises within a finite interval
of wave numbers that extends from the most unstable zero wave number. Obtained from
the long-wavelength expansion, $Ra_{c}$ at $k=0$ is not affected by the along-slot
gravity component at all. However, the along-slot gravity component still has a
stabilizing effect on small-amplitude steady perturbations with $k>0$. This
effect is manifested in that the interval of unstable wave numbers
shrinks to the vicinity of $k=0$ and vanishes with
$\theta$ decreasing from $\pi$ to $\pi/2$.

The effect of the along-slot
gravity component still gives rise to finite-amplitude
steady convection. In contrast to the finite-amplitude pattern of
symmetric counter-rotating convective cells at $\theta=\pi$, the convection
pattern at $\theta=\pi/2$ comprises only clockwise-rotating cells. Underlain
by differential gradient diffusion, the basic nature of finite-amplitude
convective steady flows at $\theta=\pi/2$ is however the same as that
at $\theta=\pi$. Continuous transformation between these flows was
thus demonstrated to take place at least up to $Ra=31000$.

In contrast to $\theta\in[\pi/2,\pi]$, finite-amplitude convective
steady flows at $\theta=0$ could not be continuously transformed into those
at $\theta=\pi/2$ for $Ra=31000$. Rather, a region of hysteresis in $\theta$
was demonstrated to develop above $Ra\approx20000$. Such hysteresis phenomenon
is a manifestation of the basic nature of finite-amplitude steady convection
at $\theta=0$ being different from that at $\theta=\pi/2$. In particular, 
differential gradient diffusion is a stabilizing factor with respect to the
effect of the across-slot gravity component on finite-amplitude steady
convection for $\theta\in[0,\pi/2)$. It is the opposite roles of this
physical process in the effects of the across-slot and along-slot
gravity components that prevents continuous transformation
between the finite-amplitude steady flows at
$\theta=0$ and $\theta=\pi/2$.

The reported mechanisms of double-component convection due to different
component boundary conditions thus comprise two basic ingredients. These
are the disparity between component stratifications caused by a convective
perturbation and the resulting feedback that maintains amplitude growth of
such a perturbation. For all slot orientations, the feedback in small-amplitude
oscillatory and steady convection is underlain by differential gradient diffusion.
In particular, such universality of the nature of small-amplitude convection is
manifested in the transformations of oscillatory marginal-stability curves into
steady ones. Finite-amplitude steady convection can however arise due to two
dissimilar feedbacks. The role of differential gradient diffusion is
destabilizing in one of them and stabilizing in the other. 

The disparity between component gradients in
perturbed state is thus the only universal feature of the
effects of different boundary conditions. This feature is however
also independent of the physical nature of the forces responsible
for the feedback. It could therefore be expected to give rise to
instabilities when the forces generated by components are other
than the buoyancy forces. Analysis of such possibilities
could provide a further insight into the effects
of different boundary conditions.

%\newpage
%\listoffigures
\newpage
\clearpage
\newpage
\clearpage
\newpage
\clearpage
\addtocounter{figure}{-1}
\newpage
\begin{figure}
%\vspace*{1cm}
\centerline{\psfig{file=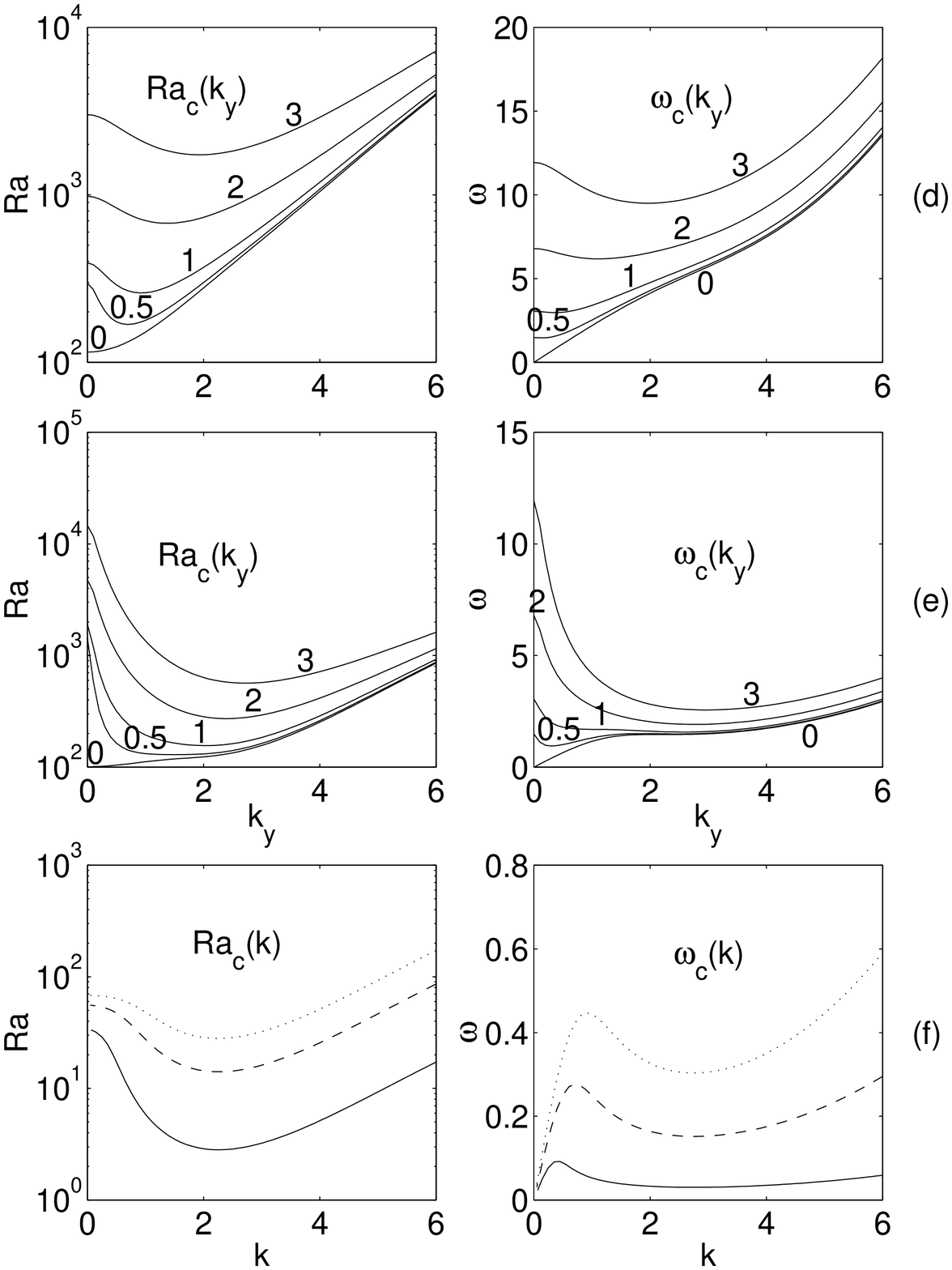,width=15cm}}
%\vspace*{0.5cm}
\caption{Inviscid fluid. Curves of the marginal linear stability to
2D ($k_{z}=0$) oscillatory disturbances, $Ra_{c}(k)$ and $\omega_{c}(k)$,
and to such 3D ($k_{z}\geq 0$) disturbances with different values of
$k_{z}$, $Ra_{c}(k_{y})$ and $\omega_{c}(k_{y})$; $\mu=1$, $Le=1$.
The numbers in (b)---(e), where the disturbances with $k_{z}>0$
were also considered, give the values of $k_{z}$ corresponding
to their nearest curve they do not intersect. (a) $\theta=0$;
(b) $\theta=\pi/8$; (c) $\theta=\pi/4$; (d) $\theta=3\pi/8$;
(e) $\theta=0.95\pi/2$; (f) the dotted lines: $\theta=0.99\pi/2$,
the dashed lines: $\theta=0.995\pi/2$, the solid lines: 
$\theta=0.999\pi/2$.}
\end{figure}
\clearpage
\newpage
\begin{figure}
\vspace*{-0.5cm}
\centerline{\psfig{file=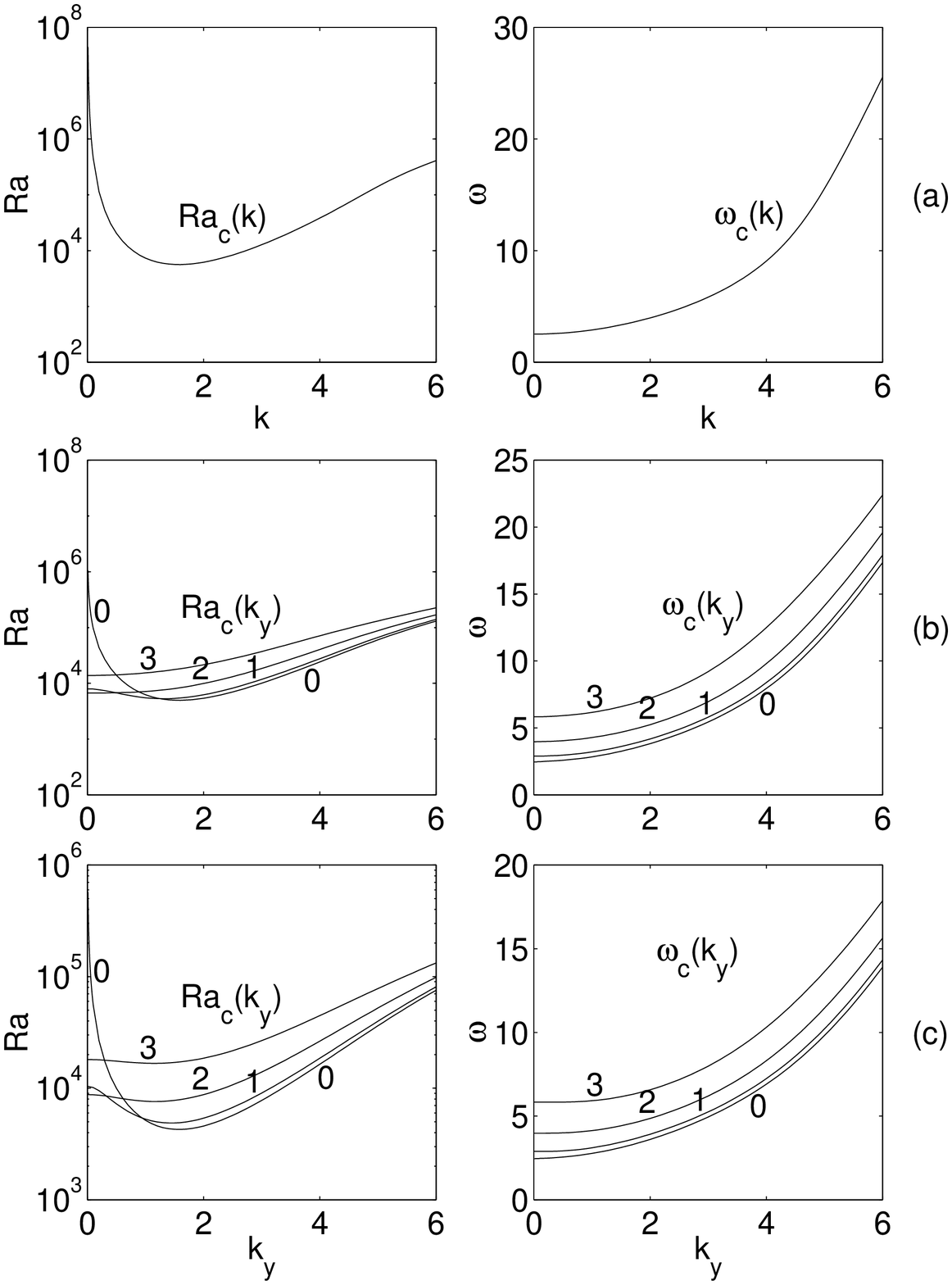,width=15cm}}
%\vspace*{0.5cm}
\caption{Viscous fluid and stress-free slot boundaries. Curves of
the marginal linear stability to 2D ($k_{z}=0$) oscillatory disturbances,
$Ra_{c}(k)$ and $\omega_{c}(k)$, and to such 3D ($k_{z}\geq 0$) disturbances
with different values of $k_{z}$, $Ra_{c}(k_{y})$ and $\omega_{c}(k_{y})$;
$\mu=1$, $Pr=6.7$, $Le=1$. The numbers in (b)---(e), where the disturbances
with $k_{z}>0$ were also considered, give the values of $k_{z}$
corresponding to their nearest curve they do not intersect.
(a) $\theta=0$; (b) $\theta=\pi/8$; (c) $\theta=\pi/4$; (d)
$\theta=3\pi/8$; (e) $\theta=\pi/2$; (f) the dash-dot lines:
$\theta=5\pi/8$, the dotted lines: $\theta=1.35\pi/2$,
the dashed lines: $\theta=1.45\pi/2$, the
solid lines: $\theta=1.49\pi/2$.}
\label{f:mscovs}
\end{figure}
\clearpage
\newpage
\addtocounter{figure}{-1}
\begin{figure}
\vspace*{-0.5cm}
\centerline{\psfig{file=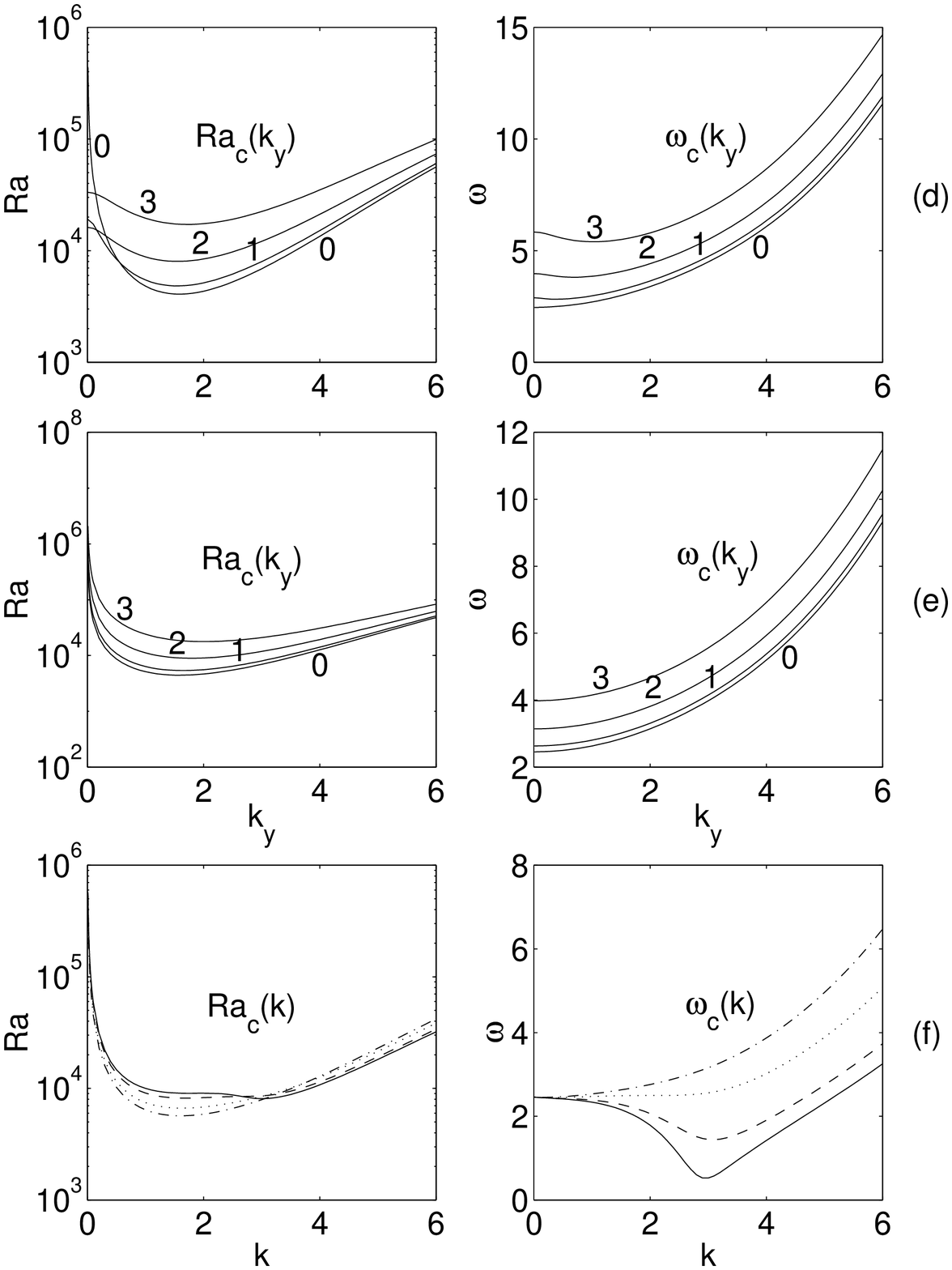,width=15cm}}
%\vspace*{0.5cm}
\caption{Viscous fluid and stress-free slot boundaries. Curves of
the marginal linear stability to 2D ($k_{z}=0$) oscillatory disturbances,
$Ra_{c}(k)$ and $\omega_{c}(k)$, and to such 3D ($k_{z}\geq 0$) disturbances
with different values of $k_{z}$, $Ra_{c}(k_{y})$ and $\omega_{c}(k_{y})$;
$\mu=1$, $Pr=6.7$, $Le=1$. The numbers in (b)---(e), where the disturbances
with $k_{z}>0$ were also considered, give the values of $k_{z}$
corresponding to their nearest curve they do not intersect.
(a) $\theta=0$; (b) $\theta=\pi/8$; (c) $\theta=\pi/4$; (d)
$\theta=3\pi/8$; (e) $\theta=\pi/2$; (f) the dash-dot lines:
$\theta=5\pi/8$, the dotted lines: $\theta=1.35\pi/2$,
the dashed lines: $\theta=1.45\pi/2$, the
solid lines: $\theta=1.49\pi/2$.}
\end{figure}
\clearpage
\newpage
\begin{figure}
\vspace*{-0.5cm}
\centerline{\psfig{file=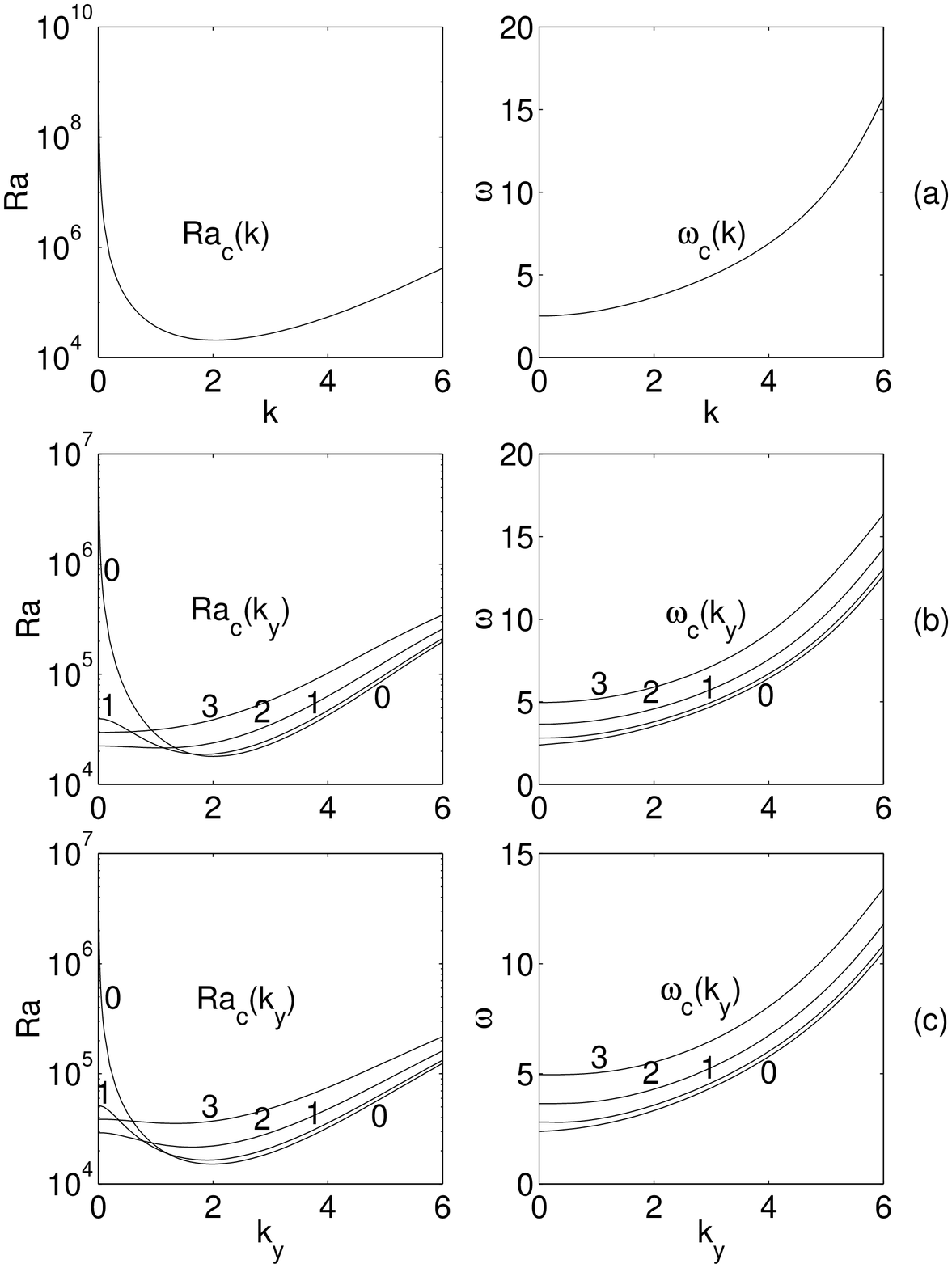,width=15cm}}
%\vspace*{0.5cm}
\caption{Viscous fluid and no-slip slot boundaries. Curves of
the marginal linear stability to 2D ($k_{z}=0$) oscillatory disturbances,
$Ra_{c}(k)$ and $\omega_{c}(k)$, and to such 3D ($k_{z}\geq 0$) disturbances
with different values of $k_{z}$, $Ra_{c}(k_{y})$ and $\omega_{c}(k_{y})$;
$\mu=1$, $Pr=6.7$, $Le=1$. The numbers in (b)---(e), where the disturbances
with $k_{z}>0$ were also considered, give the values of $k_{z}$
corresponding to their nearest curve they do not intersect.
(a) $\theta=0$; (b) $\theta=\pi/8$; (c) $\theta=\pi/4$; (d)
$\theta=3\pi/8$; (e) $\theta=\pi/2$; (f) the dash-dot lines:
$\theta=5\pi/8$, the dotted lines: $\theta=1.3\pi/2$, 
the dashed lines: $\theta=1.4\pi/2$, the
solid lines: $\theta=1.48\pi/2$.}
\label{f:mscovn}
\end{figure}
\clearpage
\addtocounter{figure}{-1}
\newpage
\begin{figure}
\vspace*{-0.5cm}
%\vspace*{1cm}
\centerline{\psfig{file=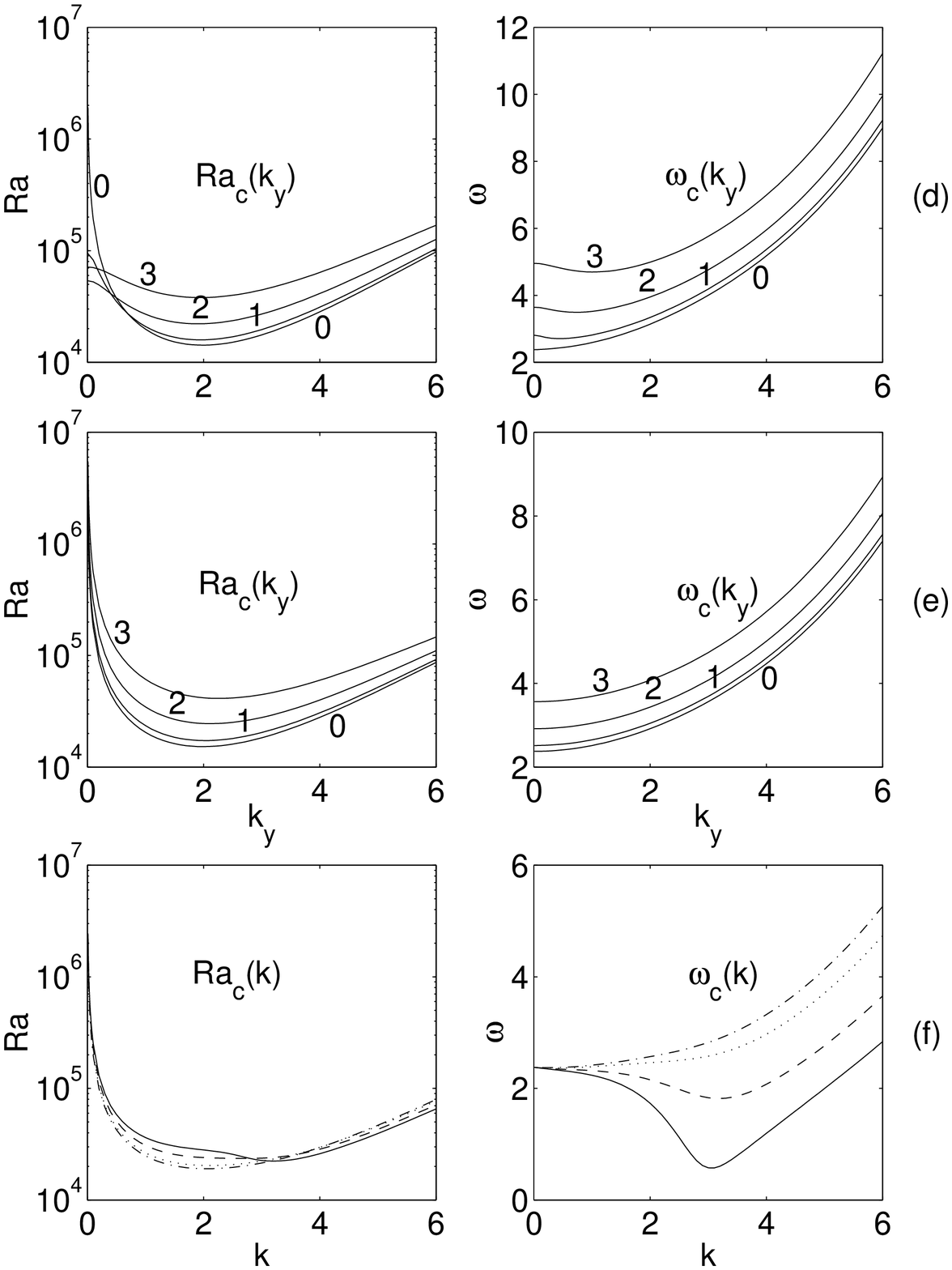,width=15cm}}
%\vspace*{0.5cm}
\caption{Viscous fluid and no-slip slot boundaries. Curves of
the marginal linear stability to 2D ($k_{z}=0$) oscillatory disturbances,
$Ra_{c}(k)$ and $\omega_{c}(k)$, and to such 3D ($k_{z}\geq 0$) disturbances
with different values of $k_{z}$, $Ra_{c}(k_{y})$ and $\omega_{c}(k_{y})$;
$\mu=1$, $Pr=6.7$, $Le=1$. The numbers in (b)---(e), where the disturbances
with $k_{z}>0$ were also considered, give the values of $k_{z}$
corresponding to their nearest curve they do not intersect.
(a) $\theta=0$; (b) $\theta=\pi/8$; (c) $\theta=\pi/4$; (d)
$\theta=3\pi/8$; (e) $\theta=\pi/2$; (f) the dash-dot lines:
$\theta=5\pi/8$, the dotted lines: $\theta=1.3\pi/2$, 
the dashed lines: $\theta=1.4\pi/2$, the
solid lines: $\theta=1.48\pi/2$.}
\end{figure}
\clearpage
\newpage
\begin{figure}
%\vspace*{1cm}
\centerline{\psfig{file=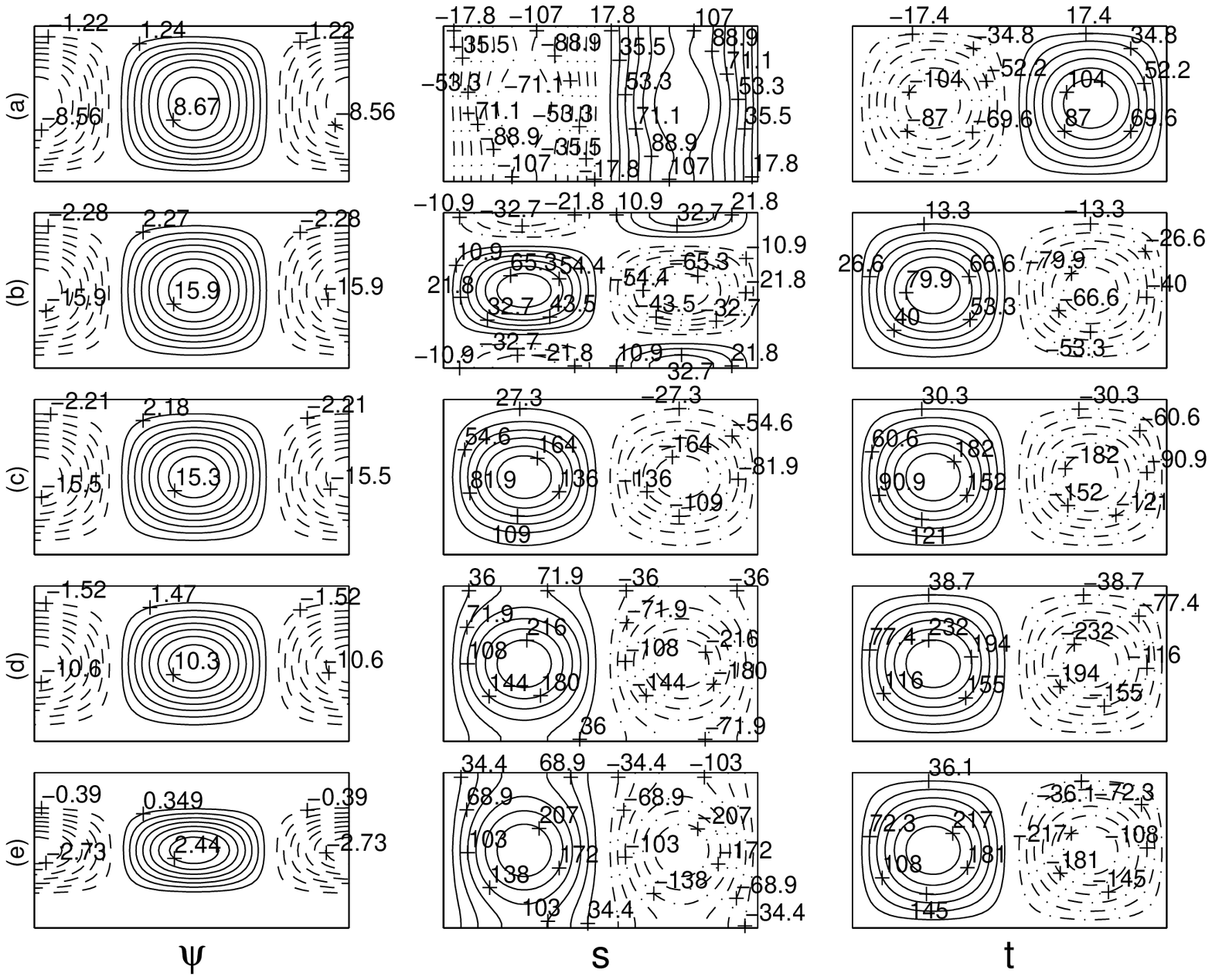,width=17cm}}
\vspace*{0.5cm}
\caption{$\theta=0$. Viscous fluid and no-slip slot boundaries. Perturbation temporal
behavior throughout the oscillation period $\tau_{p}\approx24\delta\tau$ ($\delta\tau=0.05$)
just beyond the onset of 2D oscillatory instability. It was obtained from the numerical simulation
of evolution of the linearized Eqs. (\ref{eq:ns1})---(\ref{eq:ds}) ($Ra_{S}=0$) in response to the
initial disturbance proportional to the background state after initial time $\tau_{i}\approx19000$
has passed; $\lambda=2$, $\mu=1$, $Ra=30060$, $Pr=6.7$, $Le=1$. With this $\tau_{i}$, all perturbation
modes other than the unstable mode ($\tau_{p}\approx24\delta\tau$) are practically negligible. $\psi$:
perturbation streamlines; $s$: isolines of solute concentration perturbation; $t$: perturbation
isotherms. The actual relative values of the streamfunction perturbation are equal to $10^{-3}$
times the respective values in the figure. The solid and dashed streamlines designate the
clockwise and counterclockwise rotation and are equally spaced within the positive and
negative streamfunction intervals, respectively. The solid and dash-dot isolines
of the component perturbations designate the positive and negative component
perturbation intervals, respectively.
(a) $\tau=\tau_{i}+3\delta\tau$; (b) $\tau=\tau_{i}+6\delta\tau$; (c)
$\tau=\tau_{i}+8\delta\tau$; (d) $\tau=\tau_{i}+10\delta\tau$; (e)
$\tau=\tau_{i}+12\delta\tau$; (f) $\tau=\tau_{i}+14\delta\tau$; (g)
$\tau=\tau_{i}+17\delta\tau$; (h) $\tau=\tau_{i}+20\delta\tau$; (i)
$\tau=\tau_{i}+22\delta\tau$; (j) $\tau=\tau_{i}+24\delta\tau$.}
\label{f:perh}
\end{figure}
\clearpage
\newpage
\addtocounter{figure}{-1}
\begin{figure}
%\vspace*{1cm}
\centerline{\psfig{file=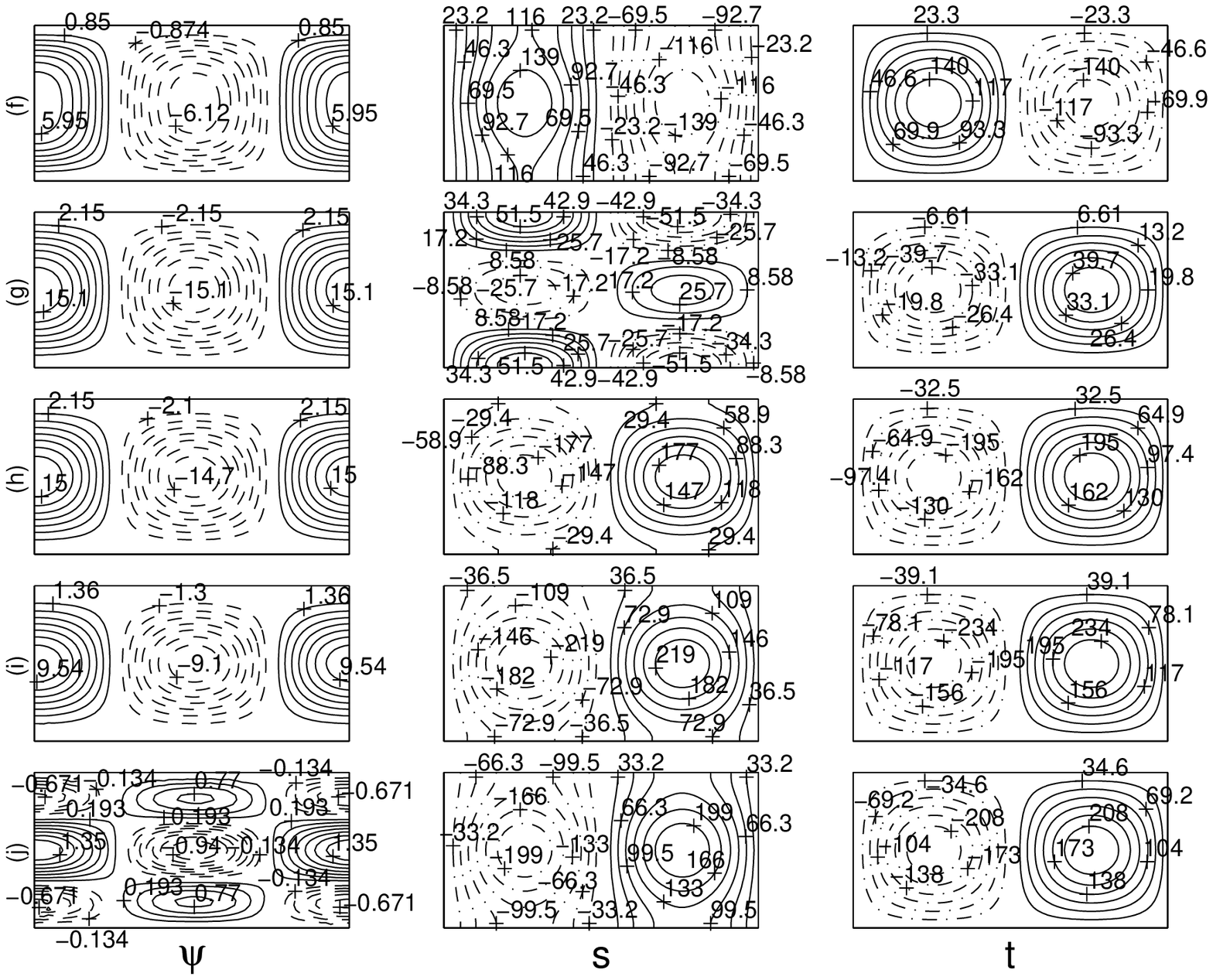,width=17cm}}
\vspace*{0.5cm}
\caption{$\theta=0$. Viscous fluid and no-slip slot boundaries. Perturbation temporal
behavior throughout the oscillation period $\tau_{p}\approx24\delta\tau$ ($\delta\tau=0.05$)
just beyond the onset of 2D oscillatory instability. It was obtained from the numerical simulation
of evolution of the linearized Eqs. (\ref{eq:ns1})---(\ref{eq:ds}) ($Ra_{S}=0$) in response to the
initial disturbance proportional to the background state after initial time $\tau_{i}\approx19000$
has passed; $\lambda=2$, $\mu=1$, $Ra=30060$, $Pr=6.7$, $Le=1$. With this $\tau_{i}$, all perturbation
modes other than the unstable mode ($\tau_{p}\approx24\delta\tau$) are practically negligible. $\psi$:
perturbation streamlines; $s$: isolines of solute concentration perturbation; $t$: perturbation
isotherms. The actual relative values of the streamfunction perturbation are equal to $10^{-3}$
times the respective values in the figure. The solid and dashed streamlines designate the
clockwise and counterclockwise rotation and are equally spaced within the positive and
negative streamfunction intervals, respectively. The solid and dash-dot isolines
of the component perturbations designate the positive and negative component
perturbation intervals, respectively.
(a) $\tau=\tau_{i}+3\delta\tau$; (b) $\tau=\tau_{i}+6\delta\tau$; (c)
$\tau=\tau_{i}+8\delta\tau$; (d) $\tau=\tau_{i}+10\delta\tau$; (e)
$\tau=\tau_{i}+12\delta\tau$; (f) $\tau=\tau_{i}+14\delta\tau$; (g)
$\tau=\tau_{i}+17\delta\tau$; (h) $\tau=\tau_{i}+20\delta\tau$; (i)
$\tau=\tau_{i}+22\delta\tau$; (j) $\tau=\tau_{i}+24\delta\tau$.}
\end{figure}
\clearpage
\newpage
\begin{figure}
\vspace*{-0.5cm}
%\vspace*{1cm}
\centerline{\psfig{file=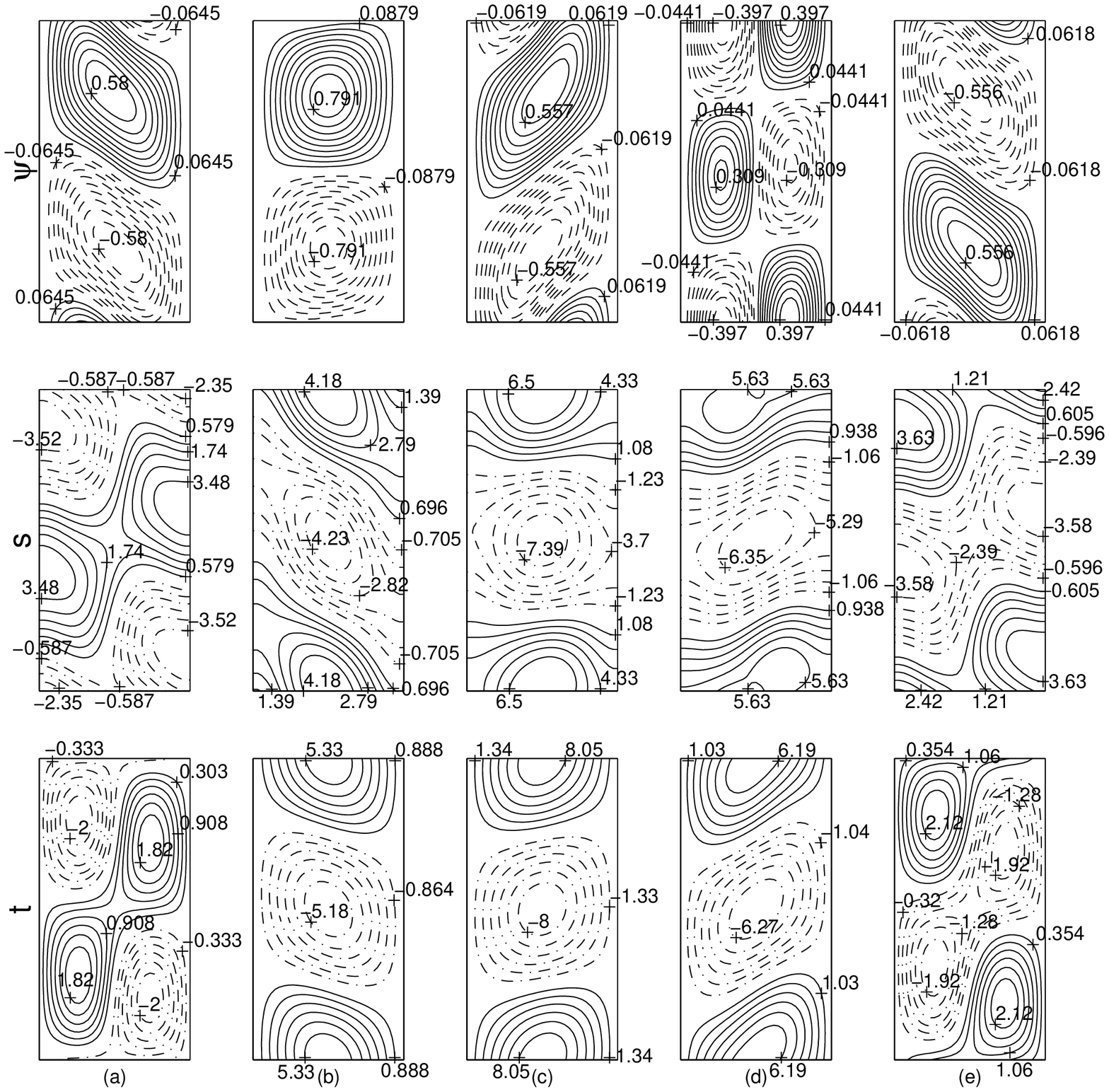,width=16.5cm}}
\vspace*{-0.3cm}
\caption{$\theta=\pi/2$. Viscous fluid and no-slip slot boundaries. Perturbation temporal
behavior throughout the oscillation period $\tau_{p}\approx34\delta\tau$ ($\delta\tau=0.05$)
just beyond the onset of 2D oscillatory instability. It was obtained from the numerical simulation
of evolution of the linearized Eqs. (\ref{eq:ns1})---(\ref{eq:ds}) ($Ra_{S}=0$) in response to the
initial disturbance proportional to the background state after initial time $\tau_{i}\approx2000$ has
passed; $\lambda=2$, $\mu=1$, $Ra=17411$, $Pr=6.7$, $Le=1$. With this $\tau_{i}$, all perturbation
modes other than the unstable mode ($\tau_{p}\approx34\delta\tau$) are practically negligible.
$\psi$: perturbation streamlines; $s$: isolines of solute concentration perturbation; $t$:
perturbation isotherms. The actual relative values of the streamfunction perturbation are
equal to $10^{-3}$ times the respective values in the figure. The solid and dashed streamlines
designate the clockwise and counterclockwise rotation and are equally spaced within the
positive and negative streamfunction intervals, respectively. The solid and dash-dot
isolines of the component perturbations are equally spaced within the positive
and negative component perturbation intervals, respectively. 
(a) $\tau=\tau_{i}+3\delta\tau$; (b) $\tau=\tau_{i}+7\delta\tau$; (c)
$\tau=\tau_{i}+11\delta\tau$; (d) $\tau=\tau_{i}+15\delta\tau$; (e)
$\tau=\tau_{i}+19\delta\tau$; (f) $\tau=\tau_{i}+22\delta\tau$; (g)
$\tau=\tau_{i}+25\delta\tau$; (h) $\tau=\tau_{i}+28\delta\tau$; (i)
$\tau=\tau_{i}+31\delta\tau$; (j) $\tau=\tau_{i}+34\delta\tau$.}
\label{f:perv}
\end{figure}
\clearpage
\newpage
\addtocounter{figure}{-1}
\begin{figure}
\vspace*{-0.5cm}
%\vspace*{1cm}
\centerline{\psfig{file=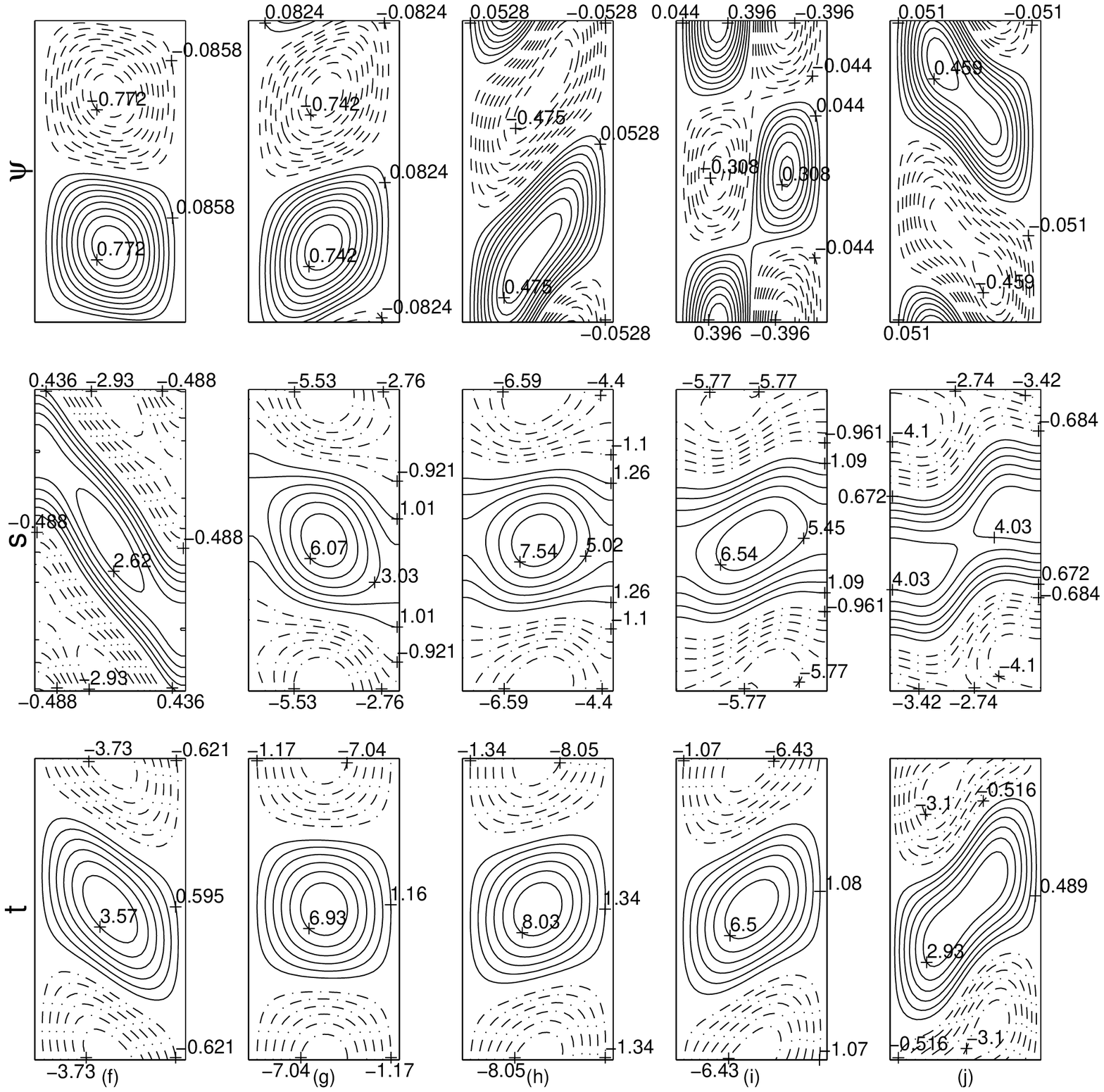,width=16.5cm}}
\vspace*{-0.3cm}
\caption{$\theta=\pi/2$. Viscous fluid and no-slip slot boundaries. Perturbation temporal
behavior throughout the oscillation period $\tau_{p}\approx34\delta\tau$ ($\delta\tau=0.05$)
just beyond the onset of 2D oscillatory instability. It was obtained from the numerical simulation
of evolution of the linearized Eqs. (\ref{eq:ns1})---(\ref{eq:ds}) ($Ra_{S}=0$) in response to the
initial disturbance proportional to the background state after initial time $\tau_{i}\approx2000$ has
passed; $\lambda=2$, $\mu=1$, $Ra=17411$, $Pr=6.7$, $Le=1$. With this $\tau_{i}$, all perturbation
modes other than the unstable mode ($\tau_{p}\approx34\delta\tau$) are practically negligible.
$\psi$: perturbation streamlines; $s$: isolines of solute concentration perturbation; $t$:
perturbation isotherms. The actual relative values of the streamfunction perturbation are
equal to $10^{-3}$ times the respective values in the figure. The solid and dashed streamlines
designate the clockwise and counterclockwise rotation and are equally spaced within the
positive and negative streamfunction intervals, respectively. The solid and dash-dot
isolines of the component perturbations are equally spaced within the positive
and negative component perturbation intervals, respectively. 
(a) $\tau=\tau_{i}+3\delta\tau$; (b) $\tau=\tau_{i}+7\delta\tau$; (c)
$\tau=\tau_{i}+11\delta\tau$; (d) $\tau=\tau_{i}+15\delta\tau$; (e)
$\tau=\tau_{i}+19\delta\tau$; (f) $\tau=\tau_{i}+22\delta\tau$; (g)
$\tau=\tau_{i}+25\delta\tau$; (h) $\tau=\tau_{i}+28\delta\tau$; (i)
$\tau=\tau_{i}+31\delta\tau$; (j) $\tau=\tau_{i}+34\delta\tau$.}
\end{figure}
\clearpage
\newpage
\begin{figure}
%\vspace*{1cm}
\centerline{\psfig{file=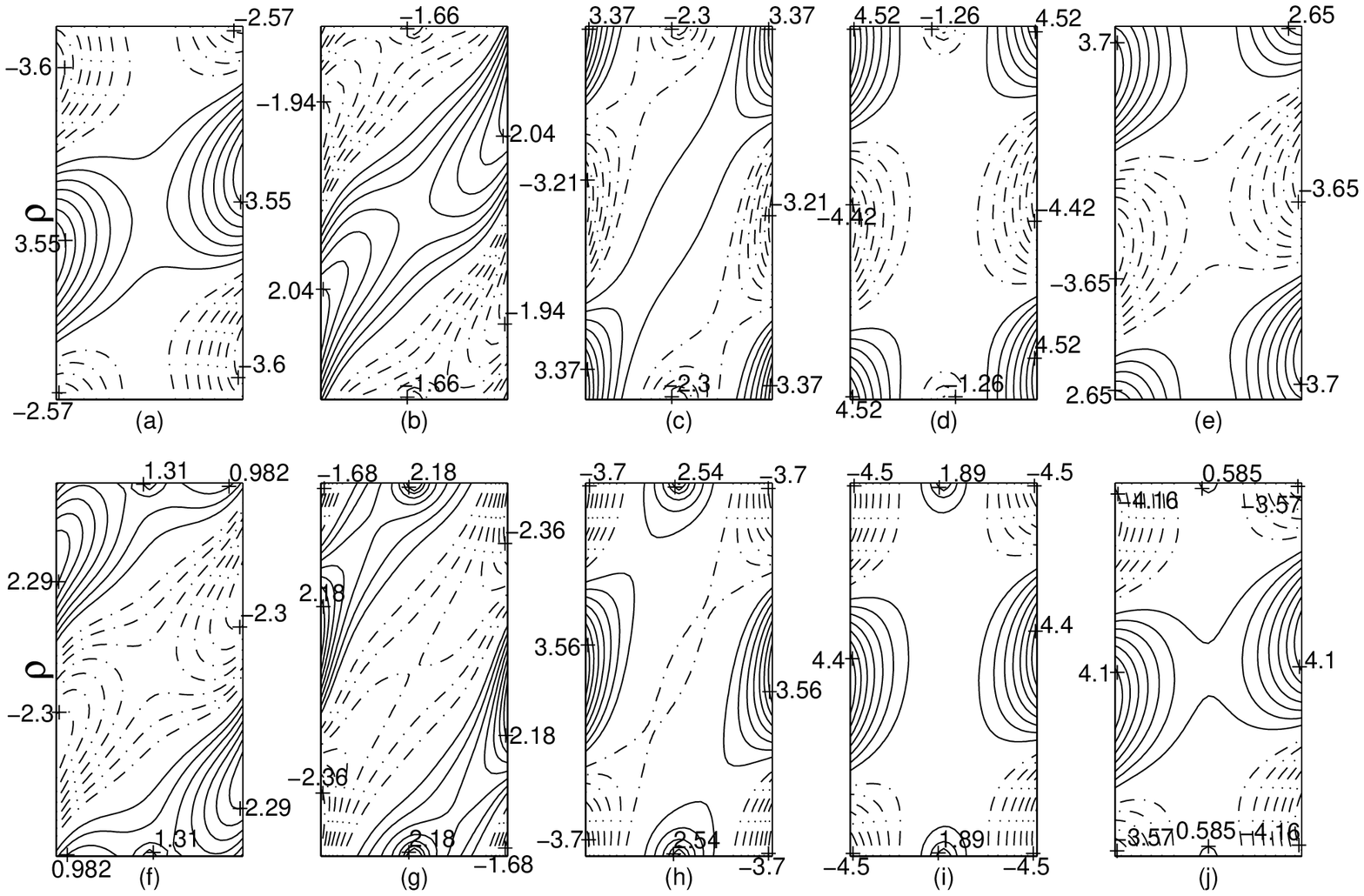,width=16.5cm}}
\vspace*{0.5cm}
\caption{The same as in Fig. \ref{f:perv}. $\rho$: isolines
of the perturbation in $s-t$. The solid and dash-dot
isolines are equally spaced within the positive
and negative density perturbation intervals,
respectively.}
\label{f:pervd}
\end{figure}
\clearpage
\newpage
\begin{figure}
%\vspace*{1cm}
\centerline{\psfig{file=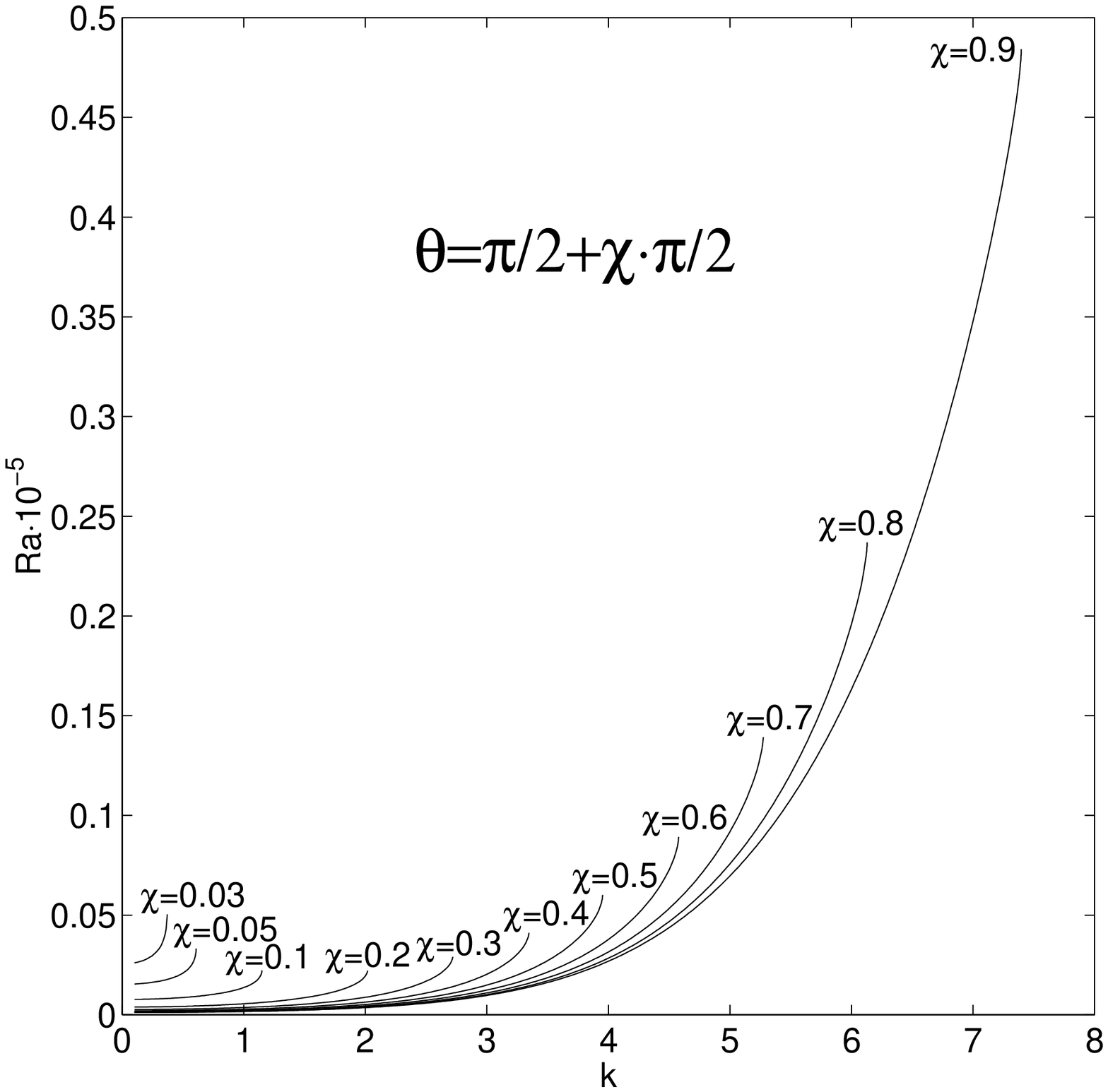,width=15cm}}
\vspace*{0.5cm}
\caption{Viscous fluid and stress-free slot boundaries.
Curves of the marginal linear stability to 2D steady disturbances,
$Ra_{c}(k)$, for different $\theta=(1+\chi)\pi/2$, $0<\chi<1$;
$\mu=1$, $Le=1$. The data for the wave numbers very close
to $k=0$ (where stable data were numerically difficult
to obtain) are not presented.}
\label{f:mscs}
\end{figure}
\clearpage
\newpage
\begin{figure}
%\vspace*{1cm}
\centerline{\psfig{file=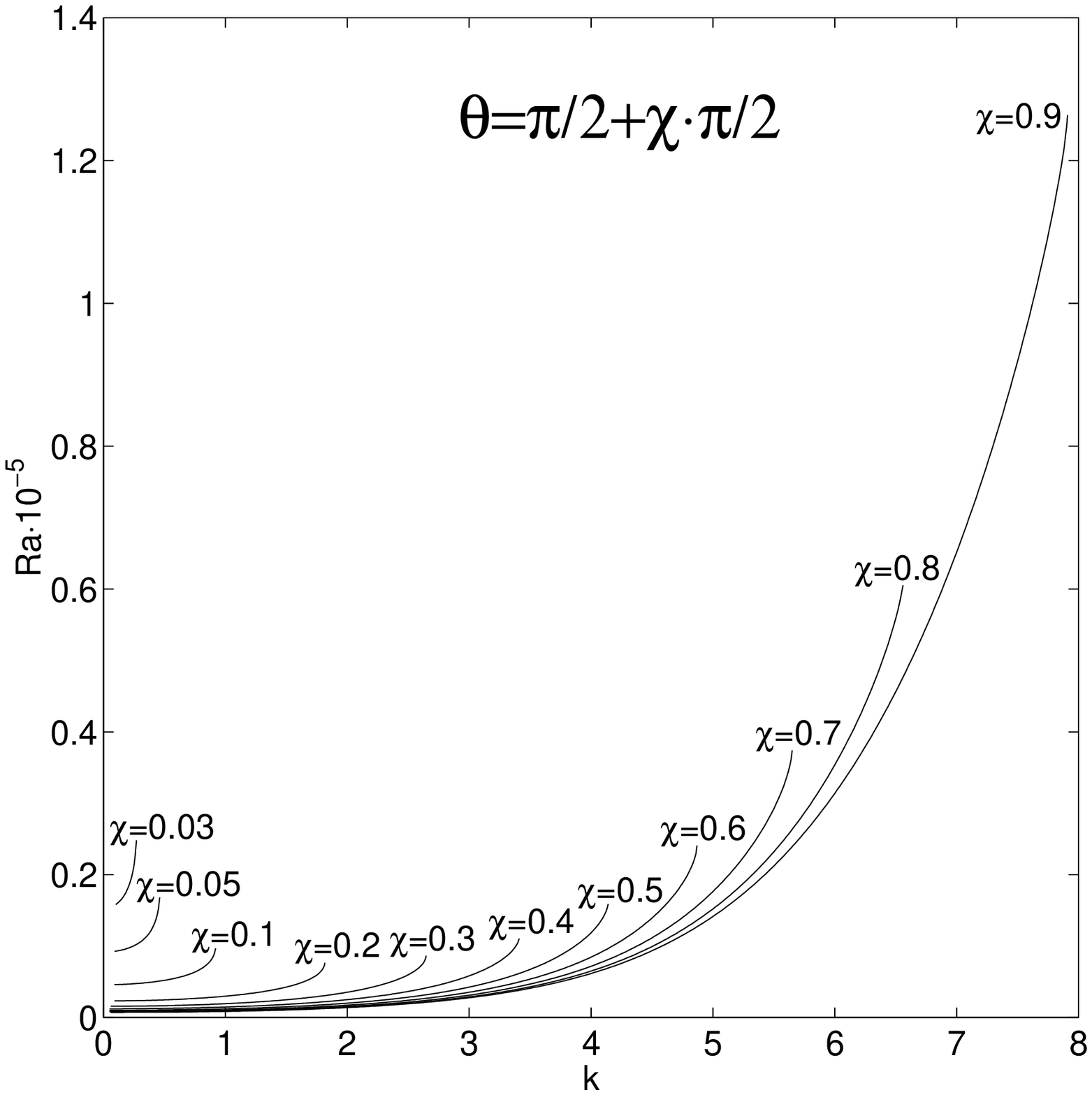,width=15cm}}
\vspace*{0.5cm}
\caption{Viscous fluid and no-slip slot boundaries.
Curves of the marginal linear stability to 2D steady disturbances,
$Ra_{c}(k)$, for different $\theta=(1+\chi)\pi/2$, $0<\chi<1$;
$\mu=1$, $Le=1$. The data for the wave numbers very close
to $k=0$ (where stable data were numerically difficult
to obtain) are not presented.}
\label{f:mscn}
\end{figure}
\clearpage
\newpage
\begin{figure}
%\vspace*{1cm}
\centerline{\psfig{file=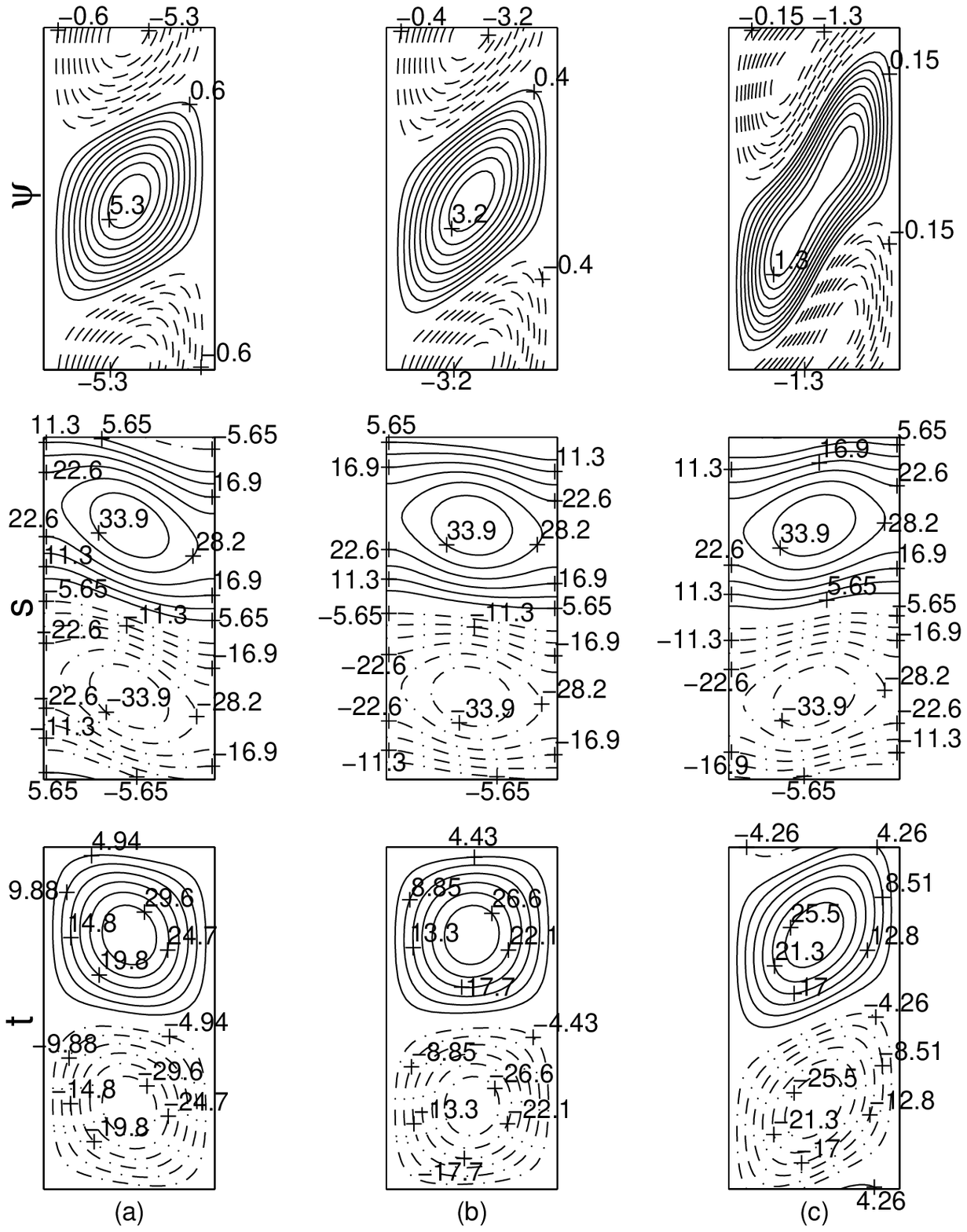,width=12.5cm}}
\vspace*{0.5cm}
\caption{$\theta=\pi/2$. Viscous fluid and no-slip slot boundaries. Singular
eigenvectors corresponding to the wavelength $\lambda=2$ at the onset of small-amplitude
steady convection within a narrow interval of $\mu$ where the linear steady instability for
$\lambda=2$ arises; $Pr=6.7$, $Le=1$. $\psi$: perturbation streamlines; $s$: isolines of
solute concentration perturbation; $t$: perturbation isotherms. The variables are 
nondimensionalized as in Eqs. (\ref{eq:ns1})---(\ref{eq:ds}) ($Ra_{S}=0$). The
actual relative values of the streamfunction perturbation are equal to $10^{-3}$
times the respective values in the figure. The solid and dashed streamlines
designate the clockwise and counterclockwise rotation and are equally
spaced within the positive and negative streamfunction perturbation intervals,
respectively. The solid and dash-dot isolines of the component
perturbations designate the positive and negative
component perturbation intervals, respectively.
(a) $\mu=0.92$, $Ra_{c}=6139$; (b) $\mu=0.96$,
$Ra_{c}=8900$; (c) $\mu=0.98$,
$Ra_{c}=22469$.}
\label{f:permu}
\end{figure}
\clearpage
\newpage
\begin{figure}
%\vspace*{1cm}
\centerline{\psfig{file=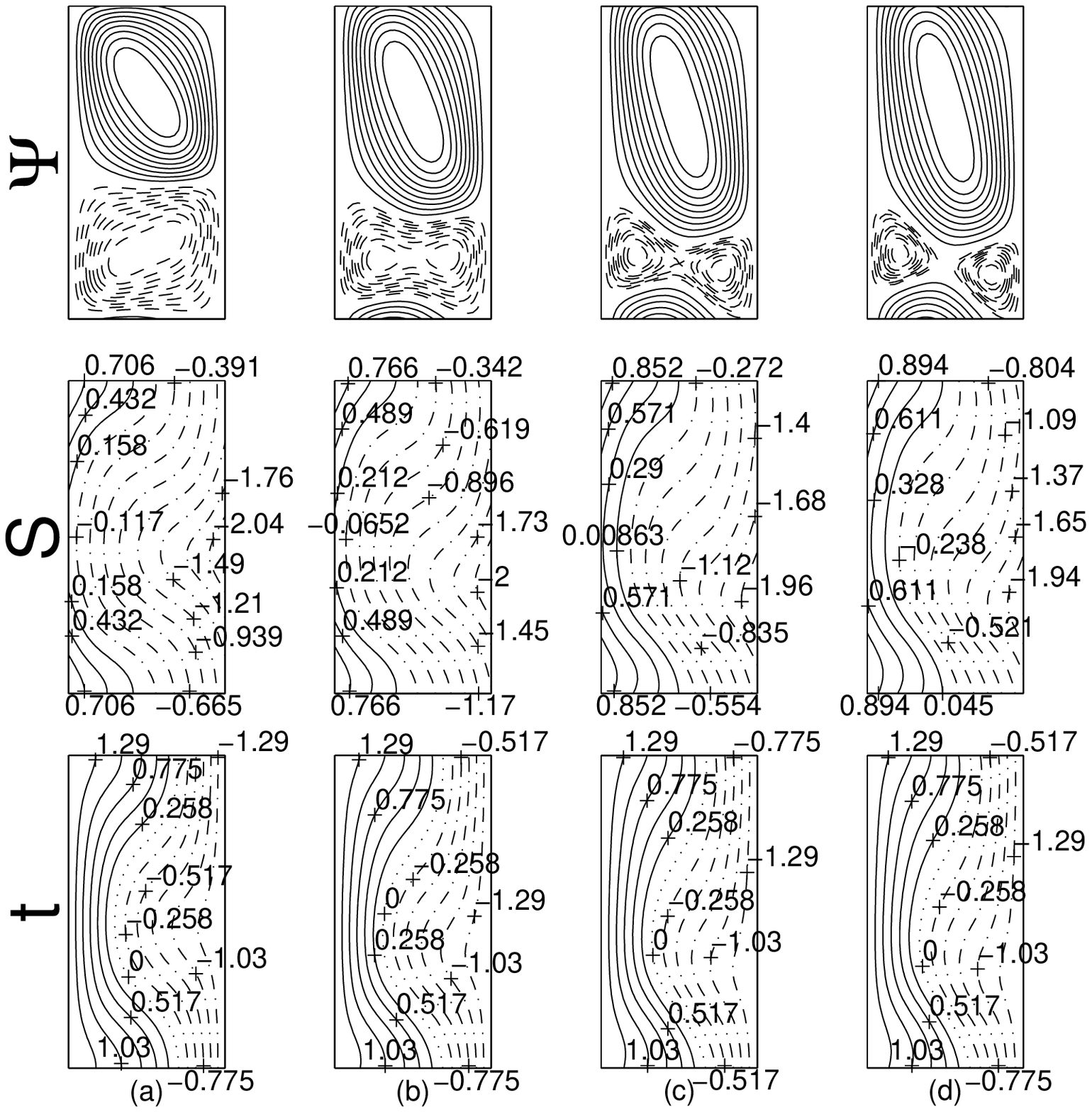,width=15cm}}
\vspace*{0.5cm}
\caption{Viscous fluid and no-slip slot boundaries. Finite-amplitude
convective steady flows representing the higher-amplitude branch, $A2$, for
different $\theta\in(\pi/2,\pi)$; $\lambda=2$, $\mu=1$, $Ra=31000$, $Pr=6.7$,
$Le=1$. The steady solutions were obtained from Eqs. (\ref{eq:ns1})---(\ref{eq:ds})
($Ra_{S}=0$). The across- and along-slot coordinate axes are directed rightwards and
upwards, respectively. $\Psi$: streamlines; $S$: isolines of solute concentration;
$t$: isotherms. The actual values of $t$ and $s$ are equal to $10^{4}$ times the
respective values in the figure. The solid and dashed streamlines designate the
clockwise and counterclockwise rotation and are equally spaced within the
positive and negative streamfunction intervals, respectively. The solid
and dash-dot component isolines designate the positive and
negative component values, respectively. The zero
isotherms are designated by the dotted lines.
(a) $\theta=1.99\pi/2$; (b) $\theta=1.97\pi/2$;
(c) $\theta=1.95\pi/2$; (d) $\theta=1.94\pi/2$;
(e) $\theta=1.9\pi/2$; (f) $\theta=1.75\pi/2$;
(g) $\theta=1.7\pi/2$; (h) $\theta=1.1\pi/2$.}
\label{f:nonsm}
\end{figure}
\clearpage
\newpage
\addtocounter{figure}{-1}
\begin{figure}
%\vspace*{1cm}
\centerline{\psfig{file=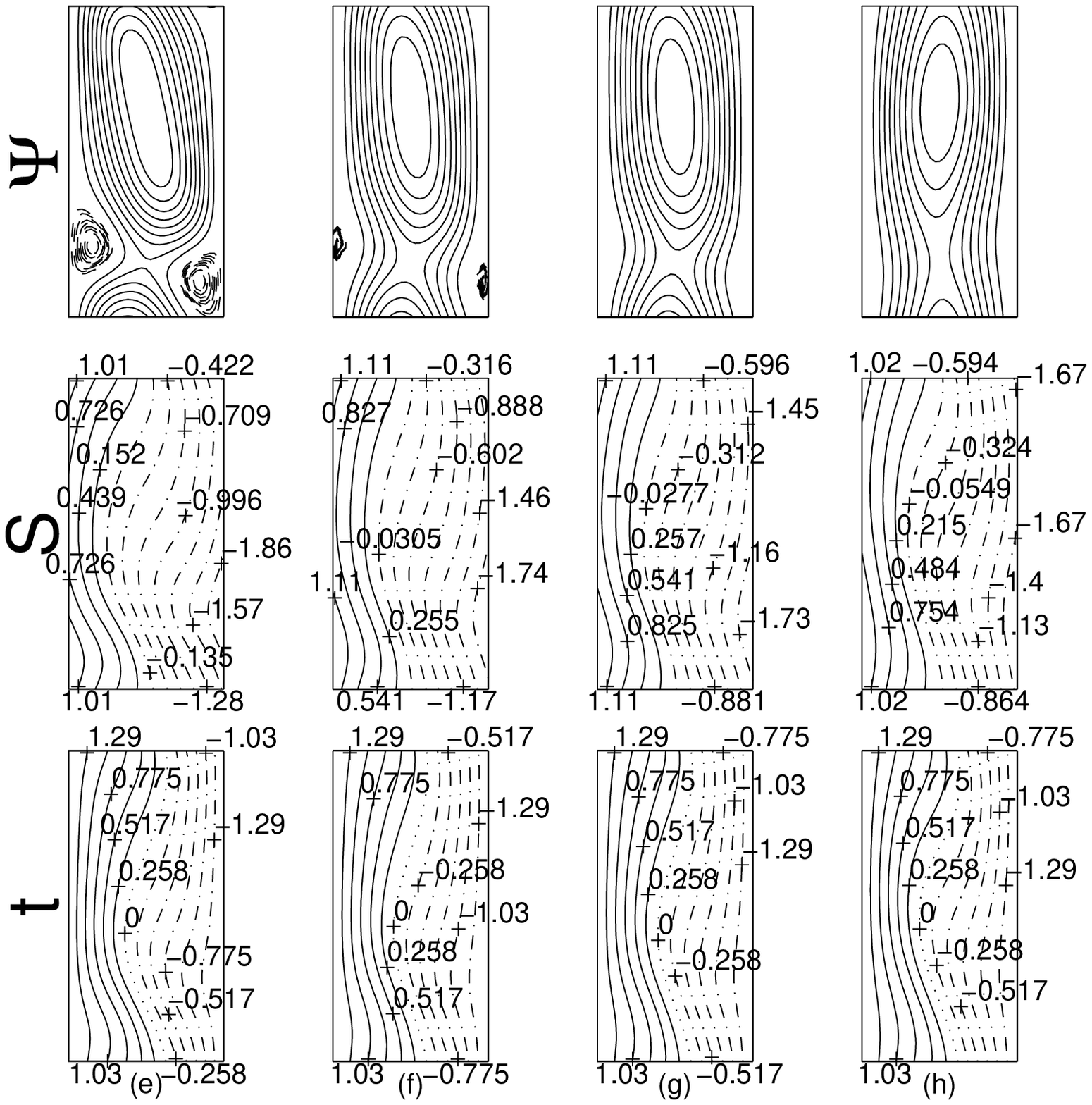,width=15cm}}
\vspace*{0.5cm}
\caption{Viscous fluid and no-slip slot boundaries. Finite-amplitude
convective steady flows representing the higher-amplitude branch, $A2$, for
different $\theta\in(\pi/2,\pi)$; $\lambda=2$, $\mu=1$, $Ra=31000$, $Pr=6.7$,
$Le=1$. The steady solutions were obtained from Eqs. (\ref{eq:ns1})---(\ref{eq:ds})
($Ra_{S}=0$). The across- and along-slot coordinate axes are directed rightwards and
upwards, respectively. $\Psi$: streamlines; $S$: isolines of solute concentration;
$t$: isotherms. The actual values of $t$ and $s$ are equal to $10^{4}$ times the
respective values in the figure. The solid and dashed streamlines designate the
clockwise and counterclockwise rotation and are equally spaced within the
positive and negative streamfunction intervals, respectively. The solid
and dash-dot component isolines designate the positive and
negative component values, respectively. The zero
isotherms are designated by the dotted lines.
(a) $\theta=1.99\pi/2$; (b) $\theta=1.97\pi/2$;
(c) $\theta=1.95\pi/2$; (d) $\theta=1.94\pi/2$;
(e) $\theta=1.9\pi/2$; (f) $\theta=1.75\pi/2$;
(g) $\theta=1.7\pi/2$; (h) $\theta=1.1\pi/2$.}
\end{figure}
\clearpage
\newpage
\begin{figure}
%\vspace*{1cm}
\centerline{\psfig{file=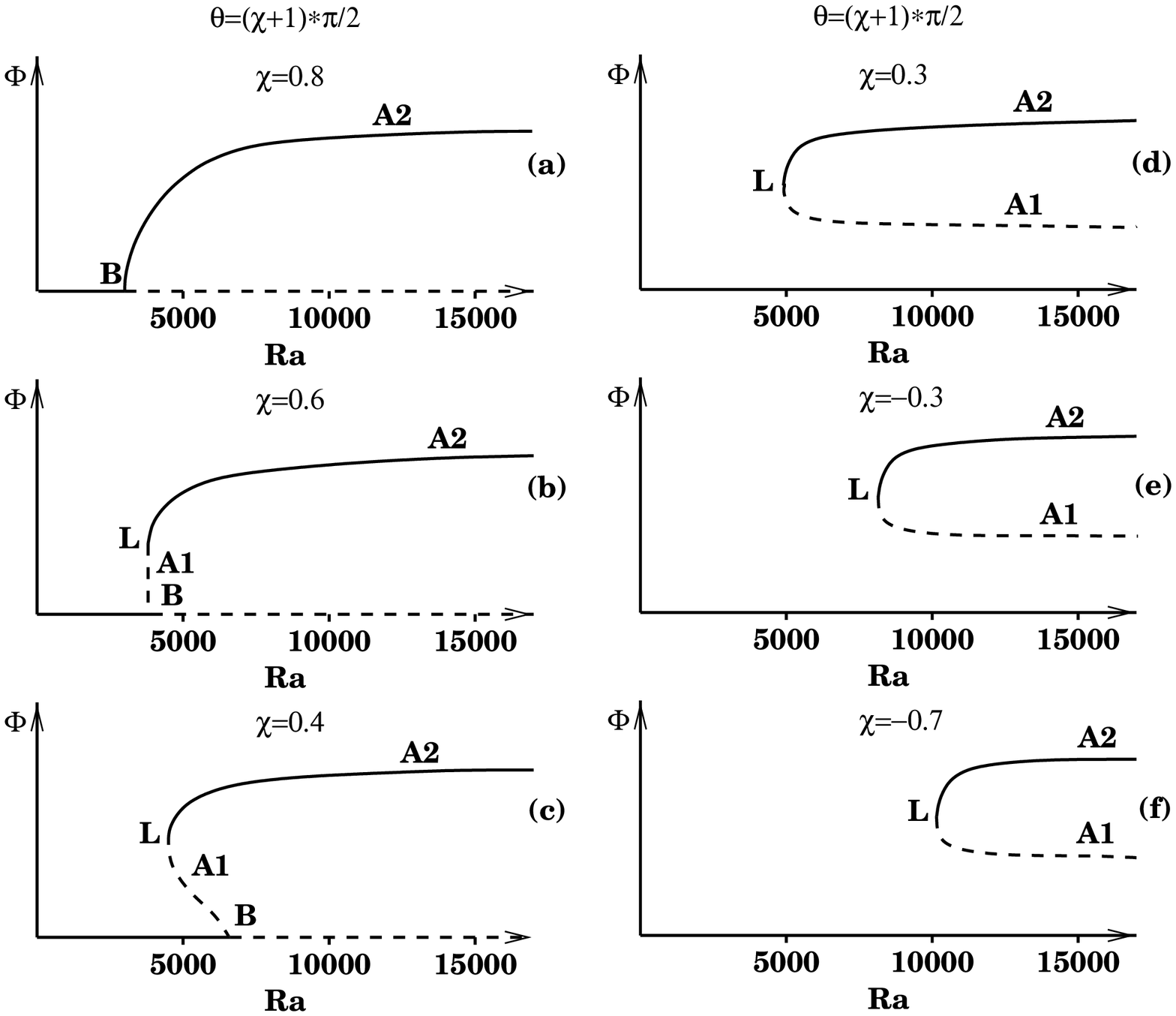,width=15cm}}
\vspace*{0.5cm}
\caption{Viscous fluid and no-slip slot boundaries. Schematic diagrams illustrating
structures of the steady flows representing onset of the linear and finite-amplitude steady
instability of the conduction state as $Ra$ is increased for different $\theta=(\chi+1)\pi/2$,
$-1<\chi<1$; $\lambda=2$, $\mu=1$, $Pr=6.7$, $Le=1$. $\Phi$ is an abstract measure of the
steady flows that distinguishes between different steady solutions, specifies the location
of the singularities (limit points and symmetry-breaking bifurcations), and represents the
flows arising from a symmetry-breaking bifurcation as a single branch. The solid (dashed) lines
represent the branches expected to be linearly stable (unstable) to the disturbances associated
with the eigenvalues that give rise to the steady instability of the conduction state. Secondary
oscillatory and steady bifurcations, if any, are not shown. $B$ is the symmetry-breaking
bifurcation standing for the steady linear stability boundary for wave number $k=\pi$
($\lambda=2$). $L$ is the limit point. $A1$ and $A2$ are the lower- and higher-amplitude
branches associated with the limit point, respectively. (a) $\theta=1.8\pi/2$;
(b) $\theta=1.6\pi/2$; (c) $\theta=1.4\pi/2$; (d) $\theta=1.3\pi/2$; (e)
$\theta=0.7\pi/2$; (f) $\theta=0.3\pi/2$.}
\label{f:bdra}
\end{figure}
\clearpage
\newpage
\begin{figure}
%\vspace*{1cm}
\centerline{\psfig{file=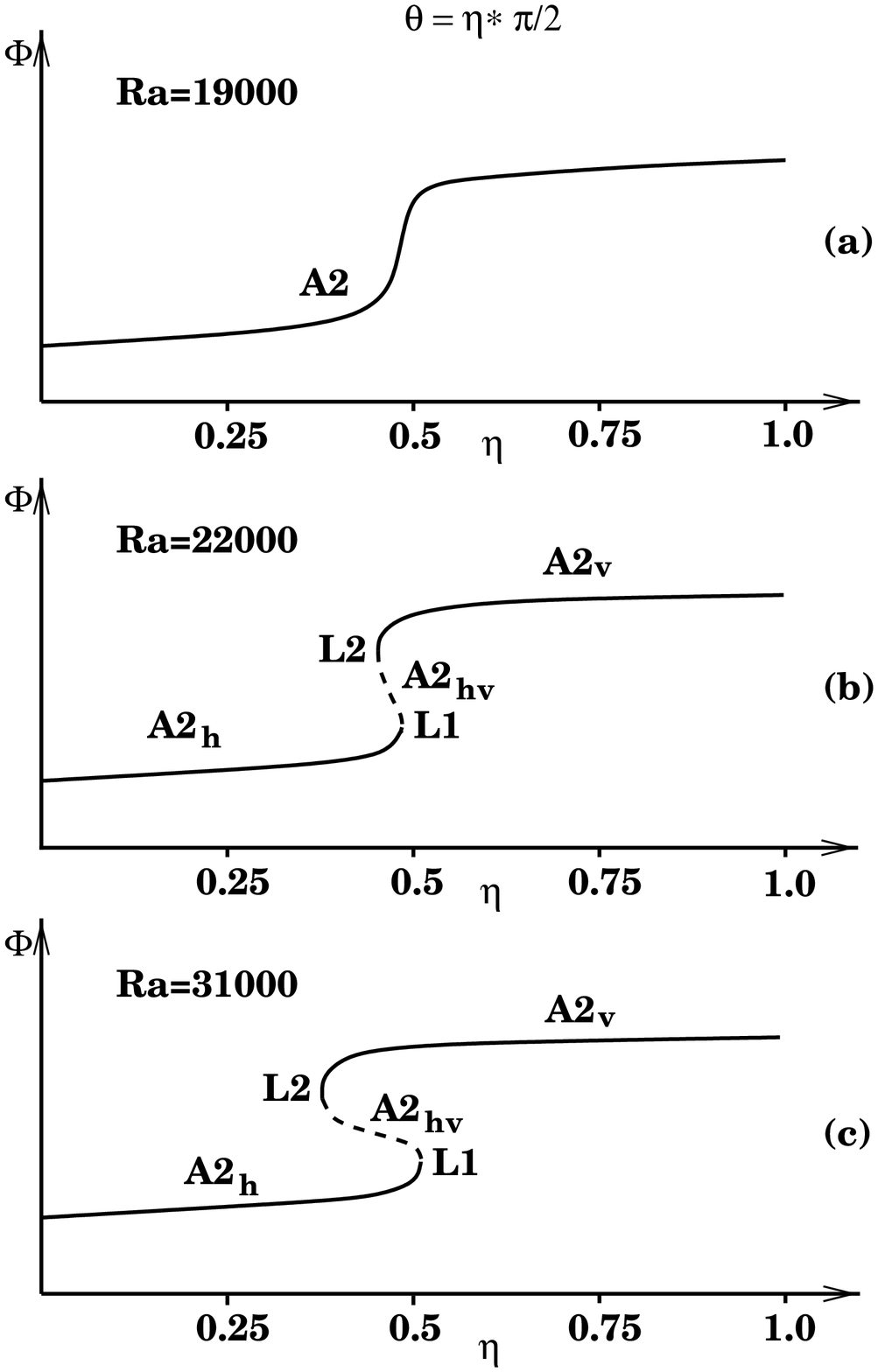,width=8.25cm}}
\vspace*{0.5cm}
\caption{Viscous fluid and no-slip slot boundaries. Schematic diagrams
illustrating structures of the steady flows of the higher-amplitude convective
branch, $A2$, with the minimal along-slot period $\lambda=2$ as $\theta=\eta\pi/2$
is varied for different $Ra$, $\eta\in[0,1]$; $\mu=1$, $Pr=6.7$, $Le=1$. $\Phi$ is
an abstract measure of the steady flows that distinguishes between different steady
solutions and specifies the location of the singularities (limit points). The dashed lines
imply that the respective branches are linearly unstable to steady disturbances. $A2_{h}$
and $A2_{v}$ are the convective branches continuously transformed into the higher-amplitude
branches in the horizontal and vertical slot, respectively. $A2_{hv}$ is the linearly unstable
(to steady disturbances) branch connecting branches $A2_{h}$ and $A2_{v}$ via limit points
$L1$ and $L2$. Secondary instability, if any, of the depicted steady branches to the
oscillatory and steady disturbances other than the (steady) disturbances
destabilizing branch $A2_{hv}$ is not shown. (a) $Ra=19000$;
(b) $Ra=22000$; (c) $Ra=31000$.}
\label{f:bdthet}
\end{figure}
\clearpage
\newpage
\begin{figure}
%\vspace*{1cm}
\centerline{\psfig{file=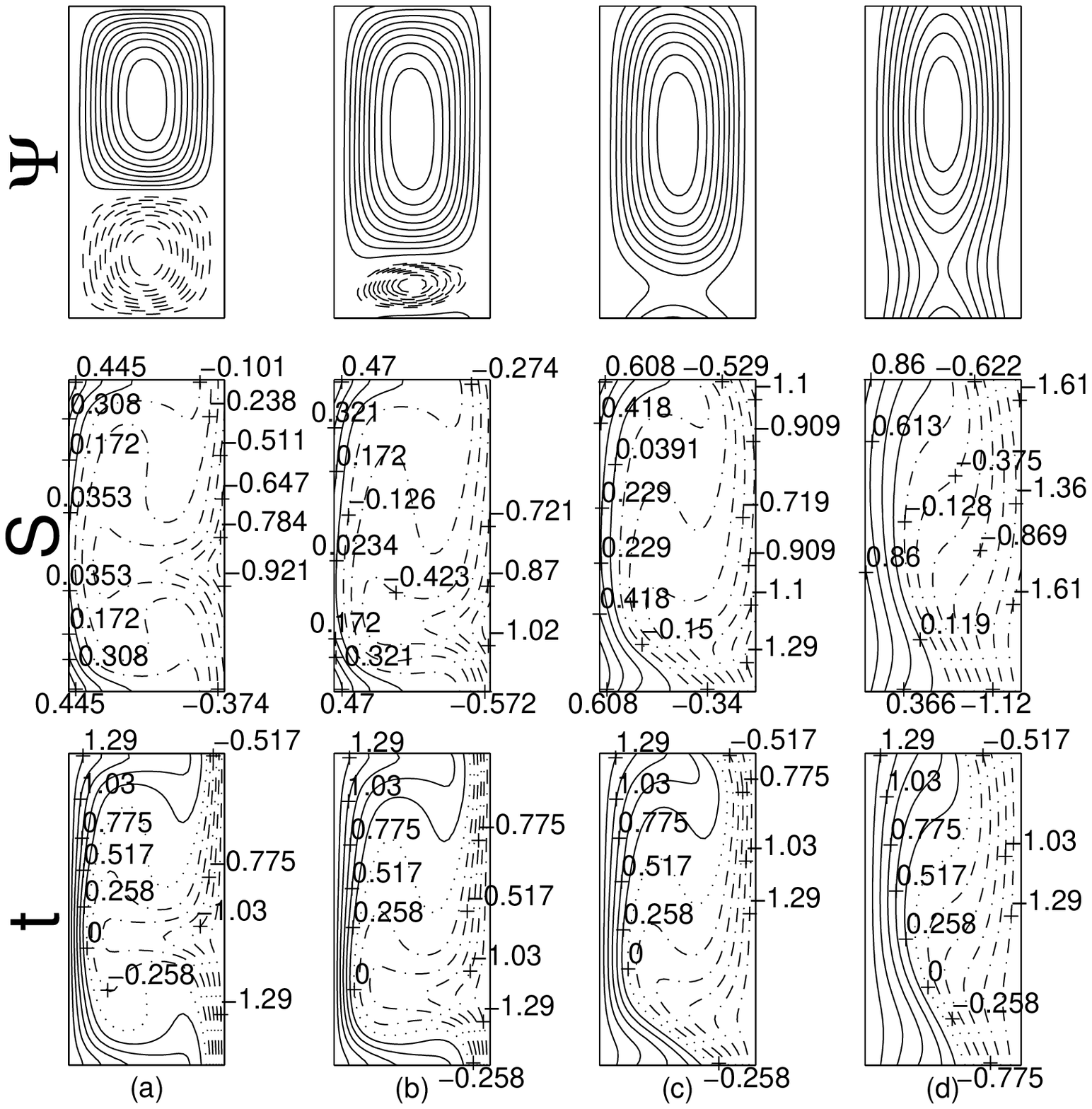,width=15cm}}
\vspace*{0.5cm}
\caption{Viscous fluid and no-slip slot boundaries. Finite-amplitude
convective steady flows representing the higher-amplitude branch, $A2$, for
different $\theta\in(0,\pi/2)$; $\lambda=2$, $\mu=1$, $Ra=31000$, $Pr=6.7$,
$Le=1$. The steady solutions were obtained from Eqs. (\ref{eq:ns1})---(\ref{eq:ds})
($Ra_{S}=0$). The across- and along-slot coordinate axes are directed rightwards
and upwards, respectively. $\Psi$: streamlines; $S$: isolines of solute concentration;
$t$: isotherms. The actual values of $t$ and $s$ are equal to $10^{4}$ times the
respective values in the figure. The solid and dashed streamlines designate the
clockwise and counterclockwise rotation and are equally spaced within the positive
and negative streamfunction intervals, respectively. The solid and dash-dot
component isolines designate the positive and negative component values,
respectively. The zero isotherms are designated by the dotted lines.
(a) $\theta=0.2\pi/2$, branch $A2_h$; 
(b) $\theta=0.5\pi/2$, branch $A2_{hv}$;
(c) $\theta=0.4\pi/2$, branch $A2_{hv}$;
(d) $\theta=0.5\pi/2$, branch $A2_v$.}
\label{f:nons}
\end{figure}
\clearpage
\newpage

\end{document}